\documentclass[fleqn,usenatbib]{mnras}

\usepackage{newtxtext,newtxmath}
\usepackage[dvipsnames]{xcolor}
\usepackage{float}
\usepackage[utf8]{inputenc}
\usepackage{graphicx}

\usepackage[T1]{fontenc}

\DeclareRobustCommand{\VAN}[3]{#2}
\let\VANthebibliography\thebibliography
\def\thebibliography{\DeclareRobustCommand{\VAN}[3]{##3}\VANthebibliography}

\usepackage{graphicx}	% Including figure files
\usepackage{amsmath}	% Advanced maths commands

\newcommand{\hsc}{\textsc{HSC~Y1}}

\title[HSC Y1 Weak Lensing Peaks and Minima]{Cosmology from weak lensing peaks and minima with Subaru Hyper Suprime-Cam survey first-year data}

\author[G.A. Marques et al.]{Gabriela A. Marques,$^{1,2,3}$\thanks{gmarques@fnal.gov}
Jia Liu,$^{4}$
Masato Shirasaki,$^{5,6}$
Leander Thiele,$^{7}$
Daniela Grandón,$^{8}$ 
\newauthor
Kevin~M.~Huffenberger,$^{3}$ 
Sihao Cheng,$^{9}$ 
Joachim Harnois-D\'eraps,$^{10}$ 
Ken Osato,$^{11,12,13}$
William R. Coulton$^{14}$
\\
$^{1}$Fermi National Accelerator Laboratory, P. O. Box 500, Batavia, IL 60510, USA\\
$^{2}$Kavli Institute for Cosmological Physics, University of Chicago, Chicago, IL 60637, USA\\
$^{3}$Florida State University, 77 Chieftan Way, Tallahassee, FL 32306, USA\\
$^{4}$Center for Data-Driven Discovery, Kavli IPMU (WPI), UTIAS, The University of Tokyo, Kashiwa, Chiba 277-8583, Japan\\
$^{5}$National Astronomical Observatory of Japan (NAOJ), National Institutes of Natural Sciences, Osawa, Mitaka, Tokyo 181-8588, Japan\\
$^{6}$The Institute of Statistical Mathematics, Tachikawa, Tokyo 190-8562, Japan\\
$^{7}$Department of Physics, Princeton University, Princeton, NJ 08544, USA\\
$^{8}$Grupo de Cosmología y Astrofísica Teórica, Departamento de Física,
FCFM, Universidad de Chile, Blanco Encalada 2008, Santiago, Chile\\
$^{9}$ Institute for Advanced Study, 1 Einstein Dr, Princeton, NJ 08540, USA\\
$^{10}$School of Mathematics, Statistics and Physics, Newcastle University, Herschel Building, NE1 7RU, Newcastle-upon-Tyne, UK\\
$^{11}$Center for Frontier Science, Chiba University, 1-33 Yayoicho, Inage, Chiba 263-8522, Japan\\
$^{12}$Department of Physics, Graduate School of Science, Chiba University, 1-33 Yayoicho, Inage, Chiba 263-8522, Japan\\
$^{13}$Kavli Institute for the Physics and Mathematics of the Universe,The University of Tokyo Institutes for Advanced Study,\\
5-1-5 Kashiwanoha, Kashiwa, Chiba 277-8583, Japan\\
$^{14}$Center for Computational Astrophysics, Flatiron Institute, 162 5th Avenue, New York, NY 10010 USA\\
}
\date{Accepted XXX. Received YYY; in original form ZZZ}

\pubyear{2023}

\begin{document}
\label{firstpage}
\pagerange{\pageref{firstpage}--\pageref{lastpage}}
\maketitle

\begin{abstract}
We present cosmological constraints derived from peak counts, minimum counts, and the angular power spectrum of the Subaru Hyper Suprime-Cam first-year (HSC Y1) weak lensing shear catalog. Weak lensing peak and minimum counts contain non-Gaussian information 
and hence are complementary to the conventional two-point statistics in constraining cosmology. In this work, we forward-model the three summary statistics and their dependence on cosmology, using a suite of $N$-body simulations tailored to the HSC Y1 data. We investigate systematic and astrophysical effects including intrinsic alignments, baryon feedback, multiplicative bias, and photometric redshift uncertainties. We mitigate the impact of these systematics by applying cuts on angular scales, smoothing scales, statistic bins, and tomographic redshift bins. By combining peaks, minima, and the power spectrum, assuming a flat-$\Lambda$CDM model, we obtain 
$S_{8} \equiv \sigma_8\sqrt{\Omega_m/0.3}= 0.810^{+0.022}_{-0.026}$, a 35\% tighter constraint than that obtained from the angular power spectrum alone. 
Our results are in agreement with other studies using HSC weak lensing shear data, as well as with Planck 2018 cosmology and recent CMB lensing constraints from the Atacama Cosmology Telescope and the South Pole Telescope.
\end{abstract}


%%%%%%%%%%%%%%%%%%%%%%%%%%%%%%%%%%%%%%%%%%%%%%%%%%

%%%%%%%%%%%%%%%%% BODY OF PAPER %%%%%%%%%%%%%%%%%%
\section{Introduction} 
\label{sec:intro}
Weak gravitational lensing (WL) is a powerful method to decode the imprints left by dark energy and dark matter in the large-scale structure (LSS). This effect arises from the coherent distortion of shapes of background galaxies by the intervening matter. The measurement of the two-point correlation function or the angular power spectrum of the weak lensing signal, known as cosmic shear, is particularly sensitive to the structure growth parameter, denoted as $S_8 \equiv \sigma_8\sqrt{\Omega_m/0.3}$. Here, $\Omega_m$ is the total matter density today, and $\sigma_8$ represents the linear matter fluctuation on a $8 h^{-1}$Mpc scale.

Recent cosmic shear results from stage-III surveys\footnote{Definition introduced by the Dark Energy Task Force report \citep{albrecht2006report}.}, such as the Kilo Degree Survey \citep[KiDS;][]{hildebrandt2020kids+,asgari2021kids,loureiro2022kids}, Dark Energy Survey \citep[DES;][]{abbott2022dark,amon2022dark,secco2022dark}, and the Subaru Hyper Suprime-Cam \citep[HSC;][]{hikage2019cosmology,hamana_hsc,dalal2023hyper,li2023hyper} have demonstrated their effectiveness in constraining cosmological information. However, the results have revealed a slight ($\approx$2$\sigma$) discrepancy between the value of $S_8$ constrained from some of these analyses and that inferred from primordial cosmic microwave background (CMB) and CMB lensing measurements \citep{planckparams,aghanim2020planck,bianchini2020constraints,qu2023atacama,dr6atacama}. Moreover, similar disagreements have been reported in some other studies using various LSS probes~\citep[e.g.,][]{aghanim2020planck,garcia2021growth,white2022cosmological, marques2023cosmological}. Additional investigations are necessary to discern if the disagreements in $S_8$ are caused by systematic effects, statistical fluctuation, or possibly new physics yet to be understood.

The weak lensing field contains non-Gaussian (NG) features resulting from the gravitational collapse of structures, structure merging, and other astrophysical processes. To fully leverage these NG features for cosmological constraints, NG statistics are required. In addition, NG statistical tools are useful for cross-validating two-point statistics for potential systematics as they often affect two-point and NG statistics differently. Various WL NG statistics have been studied in the past, such as moments~\citep{petri_mf_moments_cfhtlens,gatti_chang}, Minkowski functionals~\citep{marques2019constraining,grewal2022minkowski}, probability distribution function~\citep{Munshi2000,Bernardeau2000,liu2019constraining,Boyle2021,thiele2023cosmological,Uhlemann2023}, three-point statistics~\citep{takada_threept,3pt_fits,3pt_fu}, and deep learning~\citep{fluri2019cosmological,fluri2022full,lu2023cosmological}. In particular, the local maxima and minima on convergence maps are associated with massive halos and emptiest regions (voids) in our universe. The study of such \textit{peaks} and \textit{minima} is highly sensitive to nonlinear structures and a complementary probe to constrain $S_8$ \citep{liu2014cosmological,liu2015cosmology,kacprzak2016cosmology,shan2018kids,martinet2018kids,coulton2020weak,harnois2021cosmic,zurcher2022dark,liu2023cosmological}.\

This work presents the cosmological constraints derived from the angular power spectrum, peak counts, and minimum counts of the HSC weak-lensing first-year data (\hsc), which is the current data from this survey that is publicly available. To model the statistics, we adopt a forward modelling approach utilizing a large set of $N$-body simulations that incorporate the properties of the \hsc\ data.  

Our paper is structured as follows. In Section~\ref{sec:data_hsc} we describe the \hsc\  data. In Section~\ref{sec:sims}, we provide a description of the simulations in the forward-modeling approach. In Section~\ref{sec:method}, we describe the method for various elements of the analysis. In Section~\ref{sec:sys_nulls} we study the impact of different astrophysical and systematic effects on our results and present null tests. Finally, we present the results and internal consistency tests in Section~\ref{sec:results}, followed by the conclusions in Section~\ref{sec:conclusions}.

\section{HSC Y1 weak lensing}
\label{sec:data_hsc}
\subsection{Shape catalog}

The HSC first-year shear catalog \citep{mandelbaum2018first} (hereafter \hsc) is based on observations taken from March 2014 to April 2016 using the Subaru Hyper Suprime-Cam in five broad-bands, $grizy$. The \hsc\ is defined with conservative cuts to select galaxies with secure shape measurements, $S/N \geq 10$ and $i < 24.5$, resulting in a sample covering 136.9~$\deg^{2}$ of the sky in 6 disjoint patches: \texttt{XMM}, \texttt{GAMA09H}, \texttt{WIDE12H}, \texttt{GAMA15H}, \texttt{VVDS}, and  \texttt{HECTOMAP}.
 
The shapes $\textbf{e}=(e_{1},e_{2})$ of the galaxies are estimated on the $i-$band coadded images using the re-Gaussianization PSF correction method \citep{hirata2003shear}. In addition, the \hsc\  catalog also provide quantities to calibrate the galaxy ellipticities and to compute the corresponding shear: the intrinsic shape root mean square per component $\epsilon_{\rm rms}$, the galaxy weight $w$, the additive bias $\textbf{c}=(c_{1},c_{2})$, and multiplicative bias $m$. More information about the estimate of the shape catalog and its associated quantities can be found in \cite{mandelbaum2018first} and \cite{mandelbaum2018weakhsc}. 

The galaxy redshifts are determined from the HSC five broadband photometry using several independent codes~\citep{tanaka2018photometric}. %A detailed investigation of HSC photo-z’s was conducted by \cite{tanaka2018photometric}, who concluded that the photo-z’s ($z_p$) are most accurate at $0.2\lesssim z_{p} \lesssim 1.5$. 
For our analysis, we restrict the source redshift range to $0.3 < z_{\rm best} <1.5$, where $z_{\rm best}$ is the best-fit photo-$z$ determined by \texttt{MLZ} code. The redshift range is chosen to fall within the range deemed to be accurate by the HSC team. In Section~\ref{subsec:photoz_sims} we check the robustness of the results when considering the photo-$z$ estimated by two additional photo-$z$ codes, the classical template fitting code (\texttt{Mizuki}) and a hybrid code combining machine learning with template fitting (\texttt{Frankenz}). We apply a tomography analysis by splitting the galaxy sample into four photo-$z$ bins, with bin edges [0.3, 0.6, 0.9, 1.2, and 1.5]. However, we exclude the highest redshift bin from our analysis due to indications of unknown systematics detected prior to unblinding. %either unmodelled effects in our mocks or systematic issues in the real data such as a mis-calibration of redshift in this bin \citep{dalal2023hyper}. This decision was made to ensure a more conservative approach to our analysis. 
Table~\ref{tab:bins_properties} summarizes the properties of the individual tomographic bins, where $N_{\rm gal}$ is the number of source galaxies. $n_{g}^{\rm eff,1}$ and $n_{g}^{\text{eff},2}$ are the effective number densities as defined in \cite{heymans2012cfhtlens} and \cite{chang2013effective}, respectively:
\begin{equation}
\label{eq:h12}
    n_{g}^{\text{eff},1} = \frac{1}{\Omega_{\rm sky}} \frac{(\sum_{i}w_{i})^2}{\sum_{i}w_{i}^2}\,,
\end{equation}
\begin{equation}
\label{eq:c13}
n_{g}^{\rm eff,2} =  \frac{1}{\Omega_{\rm sky}} \sum_{i}\frac{e^2_{\rm rms}}{\sigma_{e,i}^2+e^2_{\rm rms,i}}\,.
\end{equation}

\begin{table}%[htpb]
\centering
\begin{tabular}{|c|c|c|c|}
\hline
\multicolumn{1}{|l|}{$z$-range} & \multicolumn{1}{l|}{$N_{\rm gal}$} & \multicolumn{1}{l|}{$n_{g}^{\rm eff,1}[\rm arcmin^{-2}]$} & \multicolumn{1}{l|}{$n_{g}^{\rm eff,2}[\rm arcmin^{-2}]$} \\ \hline
 $0.3<z<0.6$                    & 2655624                                                        & 5.14                                            & 4.92                                            \\
$0.6<z<0.9$                    & 2700252                                                         & 5.23                                            & 4.93                                            \\
$0.9<z<1.2$                    & 2032034                                                         & 3.99                                            & 3.61                                            \\
 \color{gray}$1.2<z<1.5$                    & \color{gray}1175188                                                    & \color{gray}2.33                                            & \color{gray}2.00                                           
  \\ \hline
\end{tabular}
\label{tab:bins_properties}
\caption{Summary of the four \hsc\ tomographic bins, each defined by the photometric redshift range $z-$range. $N_{\rm gal}$ is the total number of source galaxies. The effective number densities $n_g^{\rm eff,1}$ and $n_g^{\rm eff,2}$ as defined in %\cite{heymans2012cfhtlens} and \cite{chang2013effective}, 
Eqs.~\eqref{eq:h12} and~\eqref{eq:c13}, respectively. The highest photo-$z$ bin (shown in grey) is removed from our analysis due to systematics.}
\end{table} 

\subsection{Reconstruction of the convergence field}
\label{sec:map_making}
The shear of each galaxy $\gamma^{\rm obs}_{\alpha}$ ($\alpha=1,2)$ is estimated from the measured ellipticity as 
\begin{equation}
    \gamma_{\alpha}^{\rm obs} = \frac{1}{1+ m_{\rm tot}} \bigg(\frac{e_{\alpha}}{2\mathcal{R}} - c_{\alpha} \bigg)\,.
\label{eq:shear_2gamma}
\end{equation}
The shear responsivity $\mathcal{R}$---the response of the average galaxy ellipticity to a small shear distortion \citep{bernstein2002shapes}---is given by 
\begin{equation}
    \mathcal{R} \equiv 1-\frac{\sum_{i}w_{i}\epsilon^2_{\rm rms,i}}{\sum_{i}w_{i}}\,,
\end{equation}
where the subscript $i$ runs over all source galaxies.
 
The total multiplicative bias $m_{\rm tot}$ includes multiplicative biases in individual galaxy's shear estimation $\langle m \rangle_{i}$ and two additional terms arising from galaxy size selection and redshift-dependent responsivity corrections, $m_{\rm sel}$ and $m_{\mathcal{R}}$, respectively~\citep{mandelbaum2018weakhsc}, 
\begin{equation}
    m_{\rm tot} \equiv \langle m \rangle_{i} + m_{\rm sel}+ m_{\mathcal{R}}\,.
\label{eq:mtot_eq}
\end{equation}
The $m_{\rm sel}$ and $m_{\mathcal{R}}$ multiplicative biases are estimated considering a weighted average over the ensemble of galaxies in each tomographic bin. Specifically, $10^2 m_{\rm sel}$=[0.86, 0.99, 0.91, 0.91] and $10^2 m_{\mathcal{R}}$=[0.0, 0.0, 1.5, 3.0] in increasing order of redshift bins~\citep{hikage2019cosmology, mandelbaum2018weakhsc}.

We create the pixelized shear maps, denoted as $\boldsymbol{\hat{\gamma}}(\boldsymbol{\theta})$, for each of six HSC fields as
\begin{equation}
\boldsymbol{\hat{\gamma}}(\boldsymbol{\theta}) = W(\boldsymbol{\theta}) \boldsymbol{\gamma}^{\rm obs}(\boldsymbol{\theta})\,.
\end{equation}
Here, $W(\boldsymbol{\theta})$ represents the survey mask, which is defined as the sum of the shear weights in each pixel. To construct the maps, we consider a regular flat grid with a pixel size of $0.88\,\mathrm{arcmin}$ and apply zero-padding beyond the boundary of the image. 

The shear maps are then smoothed using the Gaussian kernel $W_{\rm G}$, defined as
\begin{equation}
      W_{\rm G}(\theta) = \frac{1}{\pi\theta_{s}^2}\exp{\bigg(-\frac{\theta^2}{\theta_{s}^2}}\bigg)\,.
    \label{eq:smooth}
\end{equation}
Initially, we consider different values for the smoothing scale $\theta_{s}$, including $1, 2, 4, 5, 8,$ and $10\,\mathrm{arcmin}$. However, to mitigate biases due to baryonic physics, we limit the analysis of peak and minimum counts to maps smoothed with $\theta_{s}= 4\,\mathrm{arcmin}$ and $\theta_{s}= 8\,\mathrm{arcmin}$ (see detailed discussions in Section~\ref{sec:baryons}). For the power spectrum, we use maps smoothed with $\theta_{s}=1\,\mathrm{arcmin}$ and apply scale cuts. In Section~\ref{sec:int_consistency}, we examine the impact of different smoothing scales on cosmological constraints.
 
We convert the shear field into a lensing convergence field following the Kaiser--Squires inversion method \citep[][KS]{kaiser1993mapping},
\begin{equation}
\tilde{\kappa}(\boldsymbol{\ell}) =   \bigg(\frac{\ell_{1}^2-\ell_{2}^2}{\ell_{1}^2+\ell_{2}^2}  \bigg)\tilde{\gamma}_{1}(\boldsymbol{\ell}) +2 \bigg(\frac{\ell_{1}\ell_{2}}{\ell_{1}^2+\ell_{2}^2} \bigg)\tilde{\gamma_{2}}(\boldsymbol{\ell})\,,
\end{equation}
where $\boldsymbol{\ell} =(\ell_{1}, \ell_{2})$ is the the multipole, and $\tilde{\kappa}$ and $\boldsymbol{\tilde{\gamma}} = (\tilde{\gamma_{1}}, \tilde{\gamma_{2}})$ are the Fourier transforms of $\kappa$ and $\boldsymbol{\hat{\gamma}}$, respectively. It is well known that missing data due to survey masks may create undesirable artifacts in the KS inversion \citep{shirasaki2013effect,liu2014mask}. To address this issue, we perform inpainting of the shear maps at the masked pixels using the sparsity property of the discrete cosine transform \citep{elad2005simultaneous,pires2009fast, Starck21} before proceeding with the KS. 

Finally, we obtained the smoothed convergence maps by conducting the inverse fast Fourier transform of $\tilde{\kappa}(\boldsymbol{\ell})$. The real part of the KS transformation is called an E-mode convergence map, which we simply refer to as the convergence field $\kappa$. The imaginary part, the so-called B-mode, refers to the divergence-free piece of the lensing field, which is zero in a noiseless scenario. As a result, B-mode will be used for null tests in Section~\ref{sec:sys_nulls}, as it may indicate systematic effects in the data. 

Following \cite{oguri2018two}, we also construct a smoothed number density map of the galaxy catalog as the convolution of the number density in each pixel with the same smoothing kernel $W_{G}$. Then, we remove the pixels where this map is less than half of the mean number density since they correspond to edges and regions that are affected by bright star masks. We re-apply the smoothed masks to the convergence maps.

\begin{figure*}
    \centering
    \includegraphics[scale=0.4]{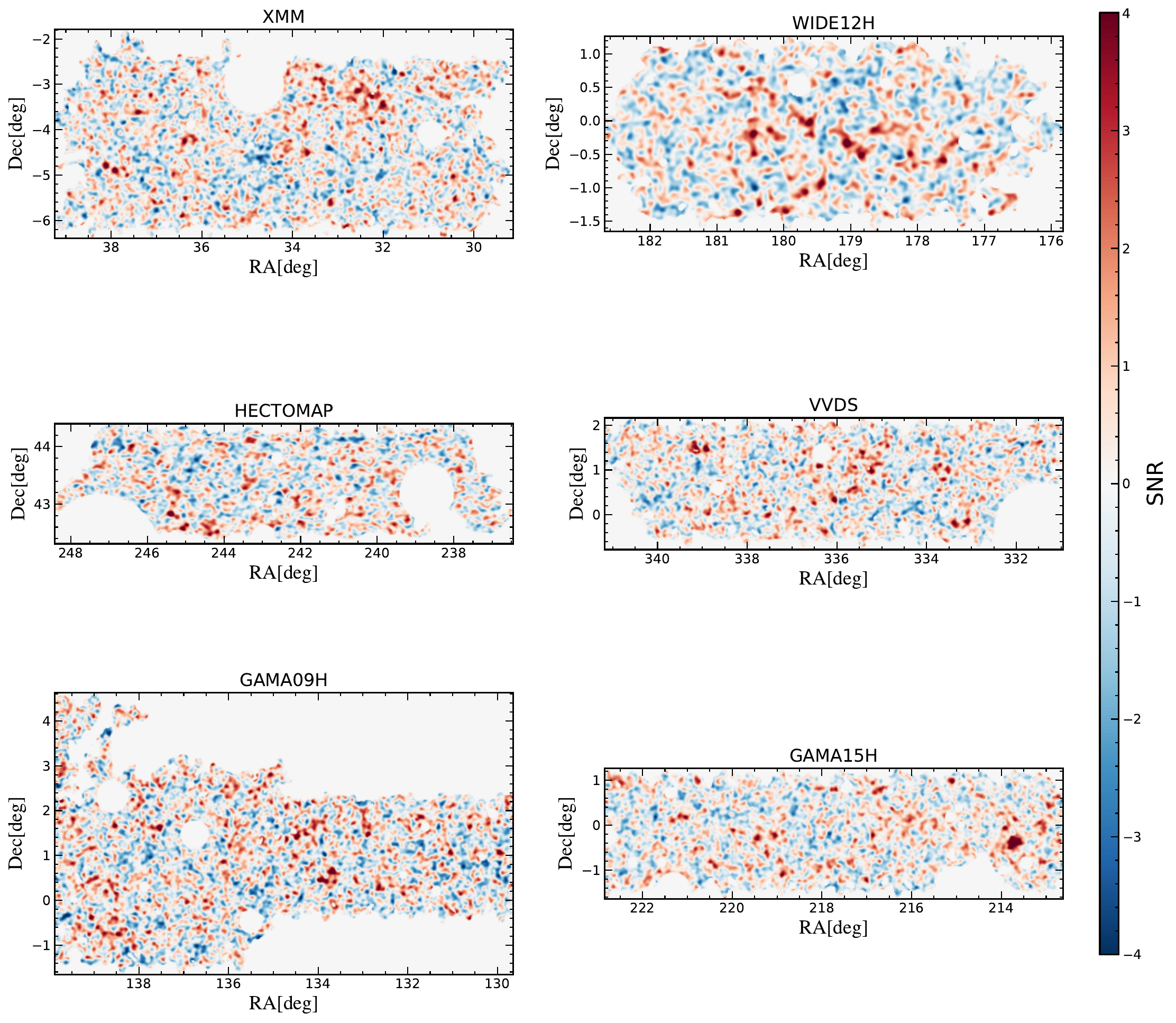}
    \caption{The reconstructed convergence fields of the six HSC Y1 regions: \texttt{XMM}, \texttt{WIDE12H}, \texttt{HECTOMAP}, \texttt{VVDS}, \texttt{GAMA09H}, and  \texttt{GAMA15H}. The maps are smoothed with a Gaussian kernel on a scale of  $2\,\mathrm{arcmin}$ for visualization purposes only. Additionally, the maps are normalized by the average standard deviation of the simulated convergence maps in fiducial cosmology. See Sections \ref{sec:data_hsc} and \ref{sec:summary_stats} for more details on the map-making and analysis choices.
}
    \label{fig:convergence_fields}
    \end{figure*}

Fig.~\ref{fig:convergence_fields} shows the weak lensing reconstructed maps for the six HSC regions. For visualization clarity, these maps are smoothed with a Gaussian kernel on a smoothing scale of $2\,\mathrm{arcmin}$. Additionally, the maps are normalized by the mean standard deviation of the convergence maps in the fiducial cosmology, as detailed in Section~\ref{sec:sims}.

\section{Simulations}
\label{sec:sims}
We rely on a large set of simulation data to build an emulator, infer cosmology, and check the effect of different systematics. Here we describe the set of simulations we use to build the covariance and emulator. We describe the simulations we use to test astrophysical and systematic effects in Section \ref{sec:sys_nulls}.

\subsection{Covariance training set}
\label{subsec:fid_sims}
We use the 2268 mock realizations of the \hsc\  shape catalogs to estimate our covariance matrix. This suite of simulations is constructed based on 108 quasi-independent full-sky $N$-body lensing simulations presented in \cite{takahashi2017full}, with the flat-$\Lambda$CDM model cosmology consistent with the best-fit result of the Wilkinson Microwave Anisotropy Probe nine-year data 
\citep{Hinshaw_2013}: $\Omega_{\text{b}}$ = 0.046, $\Omega_{\text{m}}$ = 0.279, $\Omega_\Lambda$ = 0.721, $h$ = 0.7, $\sigma_8$ = 0.82, and $n_s$ = 0.97. 
The mock is built to account for observational effects in the \hsc\ catalog, including the survey geometry, the spatial inhomogeneity of source galaxies, variations in the lensing weight, statistical noise in galaxy shape measurements, multiplicative bias, the redshifts, and distribution of the HSC galaxies. To do this, we follow the approach developed in 
\cite{Shirasaki_2014,hscmock_shirasaki}. 

The mock production can be summarized in four main steps using the information of the full-sky mocks and the observed photometric redshift, weights, shape and angular position of real
galaxies:
\begin{enumerate}
    \item The hypothetical angular coordinates (RA and DEC) of the survey window are assigned in the full-sky mock realizations. For each of the 108 full-sky simulations, it is possible to select 21 realizations of the 6 distinct HSC fields, totalizing 2268 mock shear catalogs.
    \item Next, each source galaxy are populated into one of the light-cone simulations, considering its original angular position and redshift. The photometric redshift information of the mocks is constructed based on \texttt{MLZ} method \citep{tanaka2018photometric}.
    \item The ellipticities of each source galaxy, $\boldsymbol{\epsilon}^{\rm obs}$, are randomly rotated so that the real lensing signal is erased. This done by computing $\boldsymbol{\epsilon}^{\rm ran} =\boldsymbol{\epsilon}^{\rm obs} e^{i\phi}$, where $\phi$ is a random number uniformly distributed between $0$ and $2\pi$. We model the intrinsic shape, $\boldsymbol{\epsilon}^{\rm int}$, and measurement error $\boldsymbol{\epsilon}^{\rm mea}$ as:
\begin{equation}
    \boldsymbol{\epsilon}^{\rm int} = \bigg(\frac{\epsilon_{\rm rms}}{\sqrt{\epsilon_{\rm rms}^2 +\sigma_{e}^2}} \bigg) \boldsymbol{\epsilon}^{\rm ran}\,,~  \boldsymbol{\epsilon}^{\rm mea} = N_{1} + i N_{2}\,,
\end{equation}
where $N_i$ is a random number drawn from a Gaussian distribution with a standard deviation of $\sigma_e$. The terms $\sigma_e$ and $\epsilon_{\rm rms}$ represent the shape measurement noise and the intrinsic shape dispersion per component, and are provided in the real \hsc\ catalog as \texttt{ishape\_hsm\_regauss\_derived\_sigma\_e} and \texttt{ishape\_hsm\_regauss\_derived\_rms\_e}, respectively \citep{mandelbaum2018first}. 

    \item The mock ellipticity $\boldsymbol{\epsilon}^{\rm mock}$ of each source galaxy is simulated by adding the lensing contribution at each foreground lens plane. Their components are computed as  
\begin{equation}
    \epsilon^{\rm mock}_{1} = \frac{\epsilon_{1}^{\rm int} + \delta_{1}+ (\delta_{2}/\delta^2)[1-(1-\delta^2)^{1/2}](\delta_{1}\epsilon_{2}^{\rm int} - \delta_{2}\epsilon_{1}^{\rm int} )}{1+\boldsymbol{\delta}\cdot \boldsymbol{\epsilon}^{\rm int}} + \epsilon_{1}^{\rm mea}\,,
    \label{eq:e1mock}
\end{equation}

\begin{equation}
    \epsilon^{\rm mock}_{2} = \frac{\epsilon_{2}^{\rm int} + \delta_{2}+ (\delta_{1}/\delta^2)[1-(1-\delta^2)^{1/2}](\delta_{2}\epsilon_{1}^{\rm int} - \delta_{1}\epsilon_{2}^{\rm int} )}{1+\boldsymbol{\delta}\cdot \boldsymbol{\epsilon}^{\rm int}} + \epsilon_{2}^{\rm mea}\,,
    \label{eq:e2mock}
\end{equation}

where $\boldsymbol{\delta} \equiv 2(1-\kappa)\boldsymbol{\gamma}/[(1-\kappa)^2+|\boldsymbol{\gamma}|^2]$, and $\boldsymbol{\gamma}$ and $\kappa$ are the simulated shear and convergence at the galaxy position of the light-cone simulation. We incorporate the multiplicative bias in the mock data by rescaling the shear values of Eqs.~\eqref{eq:e1mock} and \eqref{eq:e2mock} with a factor of $\boldsymbol{\gamma} \rightarrow (1 + m_{\rm tot})\boldsymbol{\gamma}$. 
\end{enumerate}
Finally, steps (ii)–(iv) are repeated for all the source galaxies.

\subsection{Cosmology varied simulations}
\label{subsec:cosmo_sims} 

To perform accurate cosmological parameter inference through a forward modelling approach, it is necessary to have a reliable prediction of the summary statistics for various cosmologies.
For that, we build a new set of \hsc\ simulations with cosmology-varied based on the simulation suite introduced in \cite{shirasaki2021noise}. These mocks consist of ray-tracing for 100 different cosmologies in the $\Omega_m-S_8$ plane, covering $0.23 \leq S_8 \leq 1.1$ and $0.1 \leq \Omega_m \leq 0.7$. The Hubble parameter is fixed at $h= 0.6727$, the baryon density $\Omega_b h^2= 0.02225$, dark energy equation-of-state parameter $w=-1$, and the spectral index $n_{\rm s}= 0.9645$.

\begin{figure}%[h]
    \centering
    \includegraphics[scale=0.4]{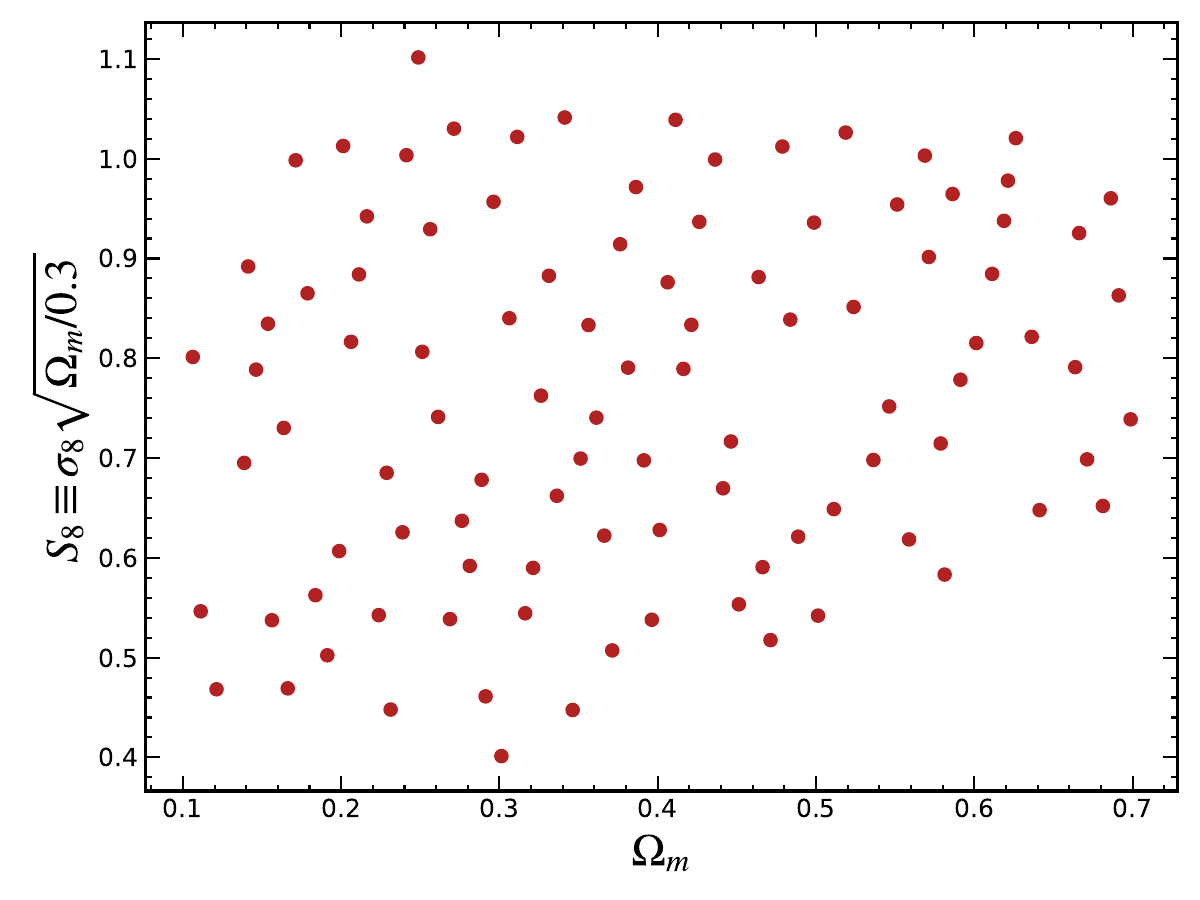}
    \label{fig:cosmo_varied_mocks}
    \caption{The distribution in the $\Omega_m$- $S_8$ plane of the simulations used to build the emulator. At each point, there are 50 mock realizations.}
\end{figure}

The $N$-body simulations were produced using the parallel Tree-Particle Mesh code \texttt{Gadget-2}, with $512^3$ particles. Varying volumes were employed to cover a broad range of redshifts, as well as have higher mass and enough spatial resolutions at lower redshifts \citep[see][]{Sato_2009, shirasaki2021noise}. A typical value of the thickness of the mass sheet is about $150$, $200$, and $300\,h^{-1}$Mpc at $z<0.5$, $0.5<z<1$, and $1<z<2$, respectively. The initial conditions were generated using a parallel code developed by \cite{Nishimichi_sims,valageas_ini}. A flat-sky approximation and the multiple lens-plane algorithm \citep{Jain_2000,hamana_takashi} were adopted to generate the ray-tracing simulations. The simulations are well converged for a wide range of multipoles and in agreement with the \texttt{halofit} prescription \citep{takahashi2012revising}. For each cosmological model, 50 ray-tracing realizations of the underlying density field were performed by randomly shifting the simulation boxes, assuming periodic boundary conditions. Finally, the cosmology-varied simulations were generated to reproduce the properties of the real data, following the steps described in Section~\ref{subsec:fid_sims}. Fig.~\ref{fig:cosmo_varied_mocks} shows the distribution of the simulations in the $\Omega_m$- $S_8$ plane. For further details on the mock production process, we refer the reader to \cite{Shirasaki_2014, shirasaki2021noise}. 
\section{Method}
\label{sec:method}
In this section, we provide an overview of the computation of summary statistics and the essential ingredients required to perform parameter inference, including emulator, covariance, and likelihood. Finally, we describe our blinding approach. 

\subsection{Summary statistics}
 \label{sec:summary_stats}

In this study, we employ three statistical measures—the angular power spectrum, peak counts, and minimum counts—to extract cosmological information from the convergence maps. 

The peaks are determined by identifying pixels that exhibit higher values compared to their eight neighboring pixels. We calculate the number of peaks within 19 equally-spaced bins, spanning the range of $-4 \leq$ SNR $\leq 4$, where SNR is defined as $\kappa/\sigma_{0}$, and $\sigma_{0}$ represents the mean standard deviation of the simulated convergence maps in fiducial cosmology. The choice of this range is based on previous studies that have demonstrated significant biases in peak estimates beyond SNR $> 4$ \citep{martinet2021impact}. 
Moreover, we do not use the information from the extreme SNR bins, where the number of peaks (or minima) is less than 15. As a result, there are 11 bins for $4,\mathrm{arcmin}$ and 8 bins for $8,\mathrm{arcmin}$ in each tomographic bin. This cut is applied to mitigate the shot-noise contribution and to use bins where the covariance is more robust. For estimating the minimum counts, we multiply the convergence maps by $-1$ and follow the same procedure as for the peak counts. As described in Sections~\ref{sec:map_making} and \ref{sec:baryons}, we use maps smoothed with Gaussian kernels of $4\,\mathrm{arcmin}$ and $8\,\mathrm{arcmin}$ to mitigate bias resulting from baryonic feedback.
 
We also utilize the 2-point information provided by the power spectrum to constrain cosmology. The lensing measurement is conducted within a complex sky mask that accounts for survey partial coverage and removes artifacts such as bright stars. However, the estimation of the angular power spectrum can be biased by the mask, leading to mode coupling between different scales. To address this issue, we adopt the Pseudo-$C_{\ell}$ approach \citep{hivon2002master} implemented by \texttt{NaMaster} \citep{alonso2019unified}, which mitigates the masking effect in the angular power spectrum estimation. Following \cite{hikage2019cosmology}, we compute the power spectrum in 14 logarithmically spaced bins, from $80 < \ell < 6500$. However, in our analysis, we use the multipole range of $300 < \ell < 1000$. The lower limit is established based on \cite{oguri2018two}, where it was found that there are unmodeled systematic errors on scales $\ell \lesssim 300$, while $\ell_{\rm max}$ is defined based on possible bias due to baryons (see Section~\ref{sec:baryons}). We do not subtract the shot-noise contribution from the $C_{\ell}^{\kappa\kappa}$ term, as the parameter inference relies on simulations that incorporate matching noise levels. We check the validity of this assumption during one of the blinding steps, as described in Section~\ref{sec:Bmode_rand}.

For each of the 3 tomographic bins, we measure the summary statistics of each individual \hsc\ field. We combine the power spectrum into a single array taking the weighted average of the 6 fields, using $\sum_{i} w_{i}$ as the weights of each field. The total peak and minimum counts are obtained by summing the results of each field.
 
\subsection{Emulator}

To model the summary statistics for a given cosmology, we construct an emulator based on the outcomes of cosmology-varied simulations. To achieve this, we utilize the Gaussian Process Regression (GPR) implemented in \texttt{scikit-learn}\footnote{\url{https://scikit-learn.org}}. Specifically, for modelling the peaks and minima, we utilize a Radial-Basis-Function with a fixed length of 6.1. For the power spectrum, we employ a Radial-Basis-Function with a fixed length of 1.0.\footnote{These values were established as optimal in the ``leave-one-out'' validation of the pipeline.} An individual Gaussian Process Regression (GPR) emulator is trained for each element of the data vectors. In Appendix~\ref{apped:emulator_valid} we assess the accuracy of the emulator using a ``leave-one-out'' validation technique. This analysis demonstrates that the emulator is capable of recovering the input statistic at a level of a few percent, which is sufficient to obtain unbiased results.

\begin{figure*}%[h]
    \centering
    \includegraphics[scale=0.6]{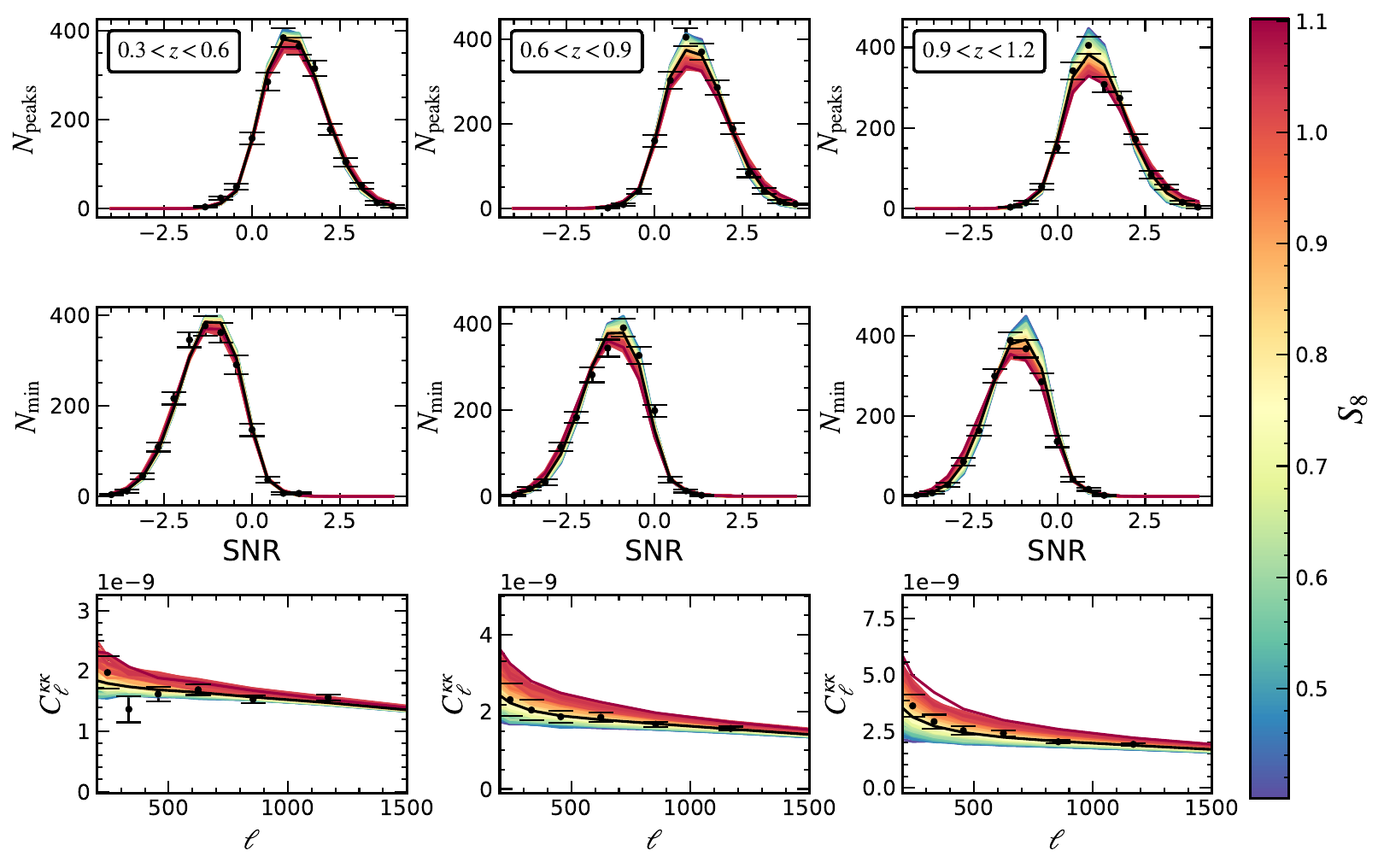}
    \caption{Peak counts (upper panel), minimum counts (middle panel) and power spectrum (lower panel) predicted for different $S_8$ values using the emulator. The black data points represent the measured statistics from the \hsc\ data, and the error bars indicate the standard deviation estimated from the simulations at the fiducial cosmology. The black line denotes the predicted statistics for the best-fit parameters. For the sake of brevity, we only present the results for a smoothing scale of  $4\,\mathrm{arcmin}$ for peaks and minima. The $8\,\mathrm{arcmin}$ smoothing is qualitatively similar.}
    \label{fig:peaks_min_stats}
\end{figure*}

Fig.~\ref{fig:peaks_min_stats} shows the sensitivity of the summary statistics for different $S_8$ values. For clarity, we display peaks (upper panel) and minimum counts (middle panel) for the maps smoothed with $4\,\mathrm{arcmin}$ only. The power spectrum of each of the tomographic bins is displayed in the lower panel.

\subsection{Covariance}

\begin{figure}%[h]
    \centering
    \includegraphics[scale=0.5]{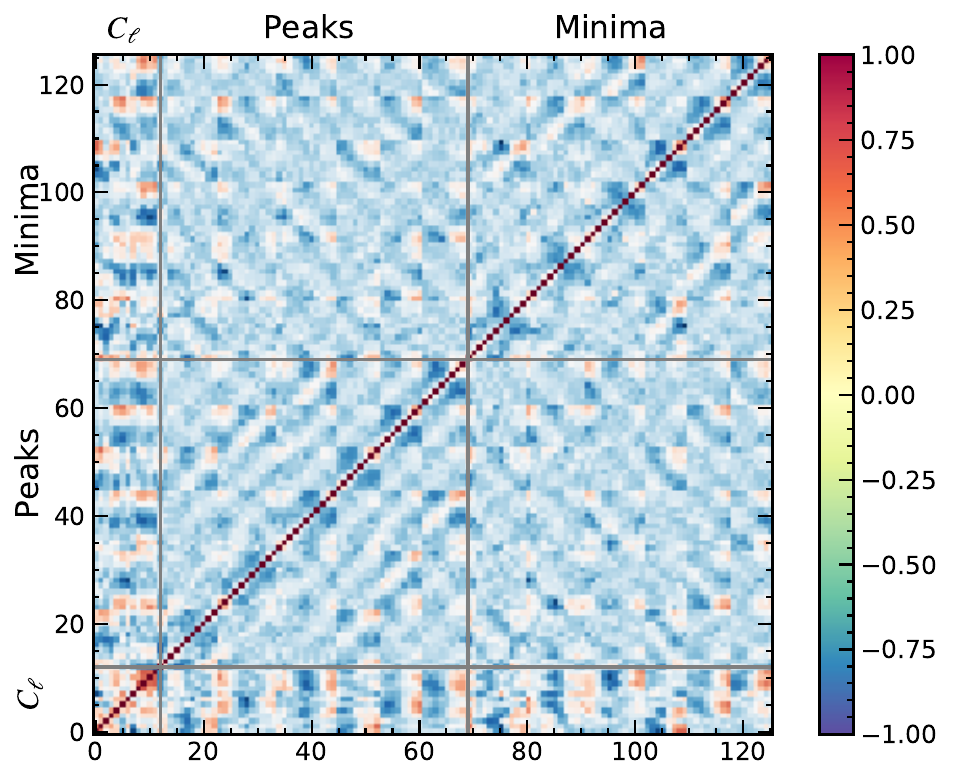}
    \caption{Correlation matrix of the peak counts, minimum counts, and power spectrum for the 3 tomographic bins. For each tomographic bin combination of the peaks and minima, we use the results of maps smoothed with $4$ and $8$ $\mathrm{arcmin}$. The off-diagonal terms showing non-trivial correlation are also included in our parameter inference.}
    \label{fig:corr_matrix}
\end{figure}

We use a set of $N_{r}= 2268$ map realizations at the fiducial cosmology to compute the covariance matrix,
\begin{equation}
    C_{ij} = \frac{1}{N_r -1} \sum_{n=1}^{N_{r}} (D^{n}_{i} - \hat{D}_{i})(D^{n}_{j} - \hat{D}_{j})\,,
\end{equation}
where $D^{n}_{i}$ is the measurement of $i-$th data component in the $n-$th sample, and $\hat{D}_{i}$ is the mean measurement of the $i-$th component. Here, the statistical descriptor can be the power spectrum, peak, minimum, or a combination of them in the case of the joint analysis. Figure~\ref{fig:corr_matrix} shows the correlation matrix between the power spectrum, peak counts, and minimum counts, for the 3 tomographic bins and smoothing scales of our analysis. We see positive (and negative) off-diagonal terms for the statistics, within their self-blocks and among them, which we include in the analysis. 

 \subsection{Data compression}
 
We implement the \texttt{MOPED} data compression to our data, to reduce the noise of the covariance matrix and Gaussianize the likelihood, particularly when dealing with multiple statistics, making the analysis more computationally efficient while preserving essential cosmological information \citep{Heavens2000,gatti_chang,zurcher2022dark}. The compressed data vector ($D^{\rm compr}$) is computed as
\begin{equation}
    D^{\rm compr} = \frac{\partial D^{\mathrm T}}{\partial p_{m}} C^{-1} D\,,
\end{equation}
where $\frac{\partial D}{\partial p_{m}}$ represents the partial derivative of the model data vector with respect to the $m$-th parameter, computed using the GPR emulator. We apply the Anderson--Hartlap factor correction \citep{hartlap2007your} when inverting the covariance matrix $C^{-1}$, to ensure an unbiased estimate. The \texttt{MOPED} procedure reduces the dimension of the uncompressed data vector $D$ to the number of model parameters considered (in our case, $\Omega_m$ and $S_8$). Consequently, the compressed data vector exhibits a more Gaussian distribution, as predicted by the central limit theorem \citep{Heavens17,gatti_chang}.

Because we discard bins outside $300< \ell<1000$, the power spectrum data vector comprises only 4 bins for each tomography bin, and its distribution is known to be nearly Gaussian. Consequently, we choose not to apply any data compression to the power spectrum. However, for peak and minimum counts, as well as when conducting a joint analysis (peak counts+ minimum counts+ power spectrum), we employ \texttt{MOPED} data compression method. We consistently apply the same data compression to the theoretical prediction when performing cosmological inference.
  
\subsection{Parameter Inference}
We perform the cosmological parameter inference, in which we evaluate the posterior of the parameters assuming a Gaussian likelihood $\mathcal{L}$ as
\begin{equation}
  \ln \mathcal{L}(\textbf{D}|\boldsymbol{\theta}) = -\frac{1}{2} [\textbf{D} - x(\boldsymbol{\theta})]^{\mathrm{T}}\mathrm{C}^{-1}[\textbf{D} - x(\boldsymbol{\theta})] 
  + \text{const.}\,,
  \label{eq:likelihood}
 \end{equation}
 where $\textbf{D}$ is the measured data vector, $x(\boldsymbol{\theta})$ is the theoretical emulator prediction at parameter values $\boldsymbol{\theta}$, and $\mathrm{C}$ is the covariance matrix computed from the fiducial simulations.
 
We sample the likelihood using the Monte Carlo Markov Chain (MCMC) sampler, implemented in the publicly available code \texttt{Cobaya}\footnote{\url{https://cobaya.readthedocs.io/en/latest/index.html}} \citep{lewis2013efficient,torrado2019cobaya}. We consider a flat prior on $0.1<\Omega_{m} < 0.4$ and $0.5<S_{8}< 1.0$, both well covered by the cosmology-varied simulations and accurate emulator. We sample the likelihood in terms of the parameter $S_8$, instead of $\sigma_8$ to obtain a uniform prior. %We determine the chain convergence using a generalized version of the $R-1$ Gelman-Rubin statistic \citep{lewis2013efficient,gelman1992inference}, which we establish that the chains converge once  $R -1 < 0.01$. 

\subsection{Blinding Strategy}
In order to avoid confirmation bias, we implement a blinding approach throughout our analysis. Since the \hsc\ data and results based on the 2-point correlation information were already unblinded in previous studies \citep{hamana_hsc, hikage2019cosmology}, we follow the steps outlined below as an honour system: 
\begin{enumerate}
    \item Pipeline development and validation: We constructed our pipeline, including map-making, emulator, covariance, and likelihood, based solely on simulations. To validate the pipeline, we used the results of the simulation as the data array and ensured that we consistently recovered the input cosmology for all the models considered within the prior.  
    \item Analysis choices: We selected scale cuts, smoothing scales, and analysis choices based on the evaluation of systematic effects, as detailed in Section~\ref{sec:sys_nulls}. Our criterion to establish the scale cuts and smoothing scales in the pipeline was chosen so that the bias introduced by systematics on $S_8$ was smaller than $0.3\sigma$, as presented in Section~\ref{sec:sys_nulls}.
    \item B-mode unblinding stage: We compared the B-mode results of the fiducial mocks with the summary statistics of the B-modes of the real data. We expect both signals to exhibit noise-like behaviour. Therefore, confirming their consistency provided further confidence that our simulations accurately represent the noise levels present in the data. See Section \ref{sec:Bmode_rand}.  
    % randomly rotated realizations of the real data. By randomly rotating the ellipticities, the lensing signal is erased, and we expect both signals to exhibit noise-like behavior. Confirming their consistency provided further confidence that our simulations accurately represent the noise levels present in the data.
    \item Power spectrum unblinding stage: We compared the power spectrum results with those presented in \cite{hikage2019cosmology} and \cite{hamana_hsc}. On the other hand, we observed a discrepancy between our results and those of the aforementioned studies when including the highest redshift bin, but consistent when using only the first three bins. A little bit after, the HSC Y3 cosmic shear analyses by \cite{dalal2023hyper, li2023hyper} were submitted, which indicated a potential bias in the highest redshift bin due to a lack of photometric redshift calibration. Consequently, we decided to exclude the highest redshift bin from our analysis to ensure the reliability and robustness of our results.
    \item Final NG statistics unblinding stage: Once we passed the previous stages, we unblinded the data vectors of peak and minimum counts and performed inferences on NG statistics as well as the joint analysis with the power spectrum.
% \end{itemize}
\end{enumerate}
Initially, our simulations only accounted for the first term of the total multiplicative bias as defined in Eq.~\eqref{eq:mtot_eq}. However, after unblinding, we discovered that, although sub-dominant, the $m_{\rm sel}$ and $m_{\mathcal{R}}$ terms needed to be incorporated in order to accurately correct the total multiplicative biases of the data~\citep{hikage2019cosmology,hamana_hsc}. Consequently, we updated our simulations to account for the complete correction and treated the data accordingly. It is important to note that this modification does not significantly impact our results (shift of $\sim 0.03\sigma$ on $S_8$), as the amplitudes of these terms are relatively small. Nonetheless, we present our results after accounting for these changes.

Before incorporating the total multiplicative bias, the smaller smoothing scales considered in the map production were $1$, $2$, and $5 \, \mathrm{arcmin}$. Given these configurations, it was observed that in order to obtain unbiased results due to baryonic feedback, for both peak and minimum counts, the minimum scale required was $5 \, \mathrm{arcmin}$. However, when producing the mocks with the total multiplicative bias, additional smoothing scales were examined, and it was found that including a smoothing scale of $4 \, \mathrm{arcmin}$ was also effective in obtaining unbiased results when appropriate cuts were applied. Considering the goal of optimizing the constraining power and capturing the non-Gaussian information of the maps, we adopt the smoothing scales of $4$ and $8\, \mathrm{arcmin}$ in our baseline setup to achieve the best possible performance in the cosmological analysis.

\section{Systematics and Null-tests}
\label{sec:sys_nulls}

In this section, we examine the impact of various systematic and astrophysical effects on our results, including photometric redshift uncertainties, image calibration uncertainties, baryonic feedback, and galaxy intrinsic alignment. To assess the impact of these effects, we conducted parameter inference for different analysis choices using distinct mocks specifically designed to simulate each effect, as described below in Sections \ref{subsec:photoz_sims}-\ref{subsec:IA}. In this process, we replaced the synthetic data vector at the fiducial cosmology with another synthetic data vector that was contaminated with each of these systematics. Our goal was to analyze the shifts in $S_8$ resulting from these substitutions, allowing us to understand the consequences of any mis-modelling of these effects. The necessary scale cuts and smoothing scales in our pipeline are defined ensuring that any shifts in $S_8$ with respect to the constraints of the fiducial, uncontaminated mocks remain below $0.3\sigma$.

\begin{figure}%[h]
\centering
    \includegraphics[scale=0.6]{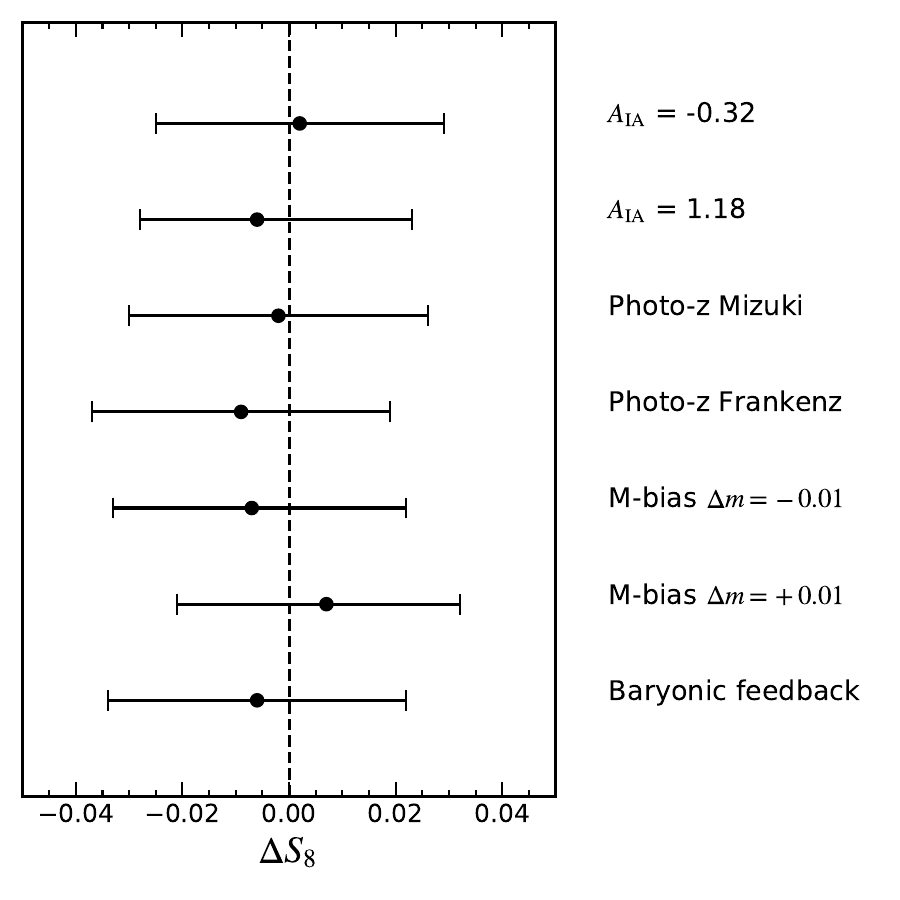}
    \caption{Impact of systematic effects on $S_8$ constraints considering the analysis choices of our baseline analysis. We use simulations to check the expected shifts on $S_8$, including baryonic feedback mechanisms, intrinsic alignments, and uncertainties of photometric redshifts and  of the multiplicative bias.}
    \label{fig:sys}
\end{figure}

Fig.~\ref{fig:sys} shows the shift on $S_8$ with respect to the constraints from the fiducial cosmology for each systematic. We show the results corresponding to our baseline setup, which includes the power spectrum with $300<\ell<1000$ and peak counts jointly with minimum counts of maps smoothed with $4 \, \mathrm{arcmin}$ and $8 \, \mathrm{arcmin}$, while also excluding extreme bins with counts less than 15.  For the baseline setup, all systematics introduce biases that are smaller than our predefined acceptable criterion for this analysis. Therefore, we proceed with the analysis by applying the necessary cuts, rather than marginalizing over these effects. This reinforces the reliability of our parameter estimation, demonstrating that the baseline setup is robust against the considered systematics. However, it is worth noting that a full error propagation of the uncertainties in shape calibration, photo-z, baryons, and IA achieved through marginalization over nuisance parameters would consistently result in larger error bars for all statistics.\

For all the effects we investigated, we observed that baryons had the most significant impact on causing shifts in $S_8$ when considering smaller scales. Specifically, for the baryonic feedback in an HSC-like survey, we found that the maps should be smoothed with smoothing scales larger than $4 \, \mathrm{arcmin}$, and we should limit the power spectrum analysis to $\ell<1000$ in order to get unbiased results according to our criterion. 
A further study of the impact of baryonic feedback using different models in these NG statistics for the HSC data is explored in \cite{Daniela2023}. %Our findings indicate that a model with weak baryonic feedback scenarios, as considered in the $\kappa$TNG simulations, is more likely to reproduce the baryonic mechanism present in \hsc~ real data or that systematic uncertainties might play a more significant role in shaping the cosmological constraints. 
 
In addition, we conduct a null test in Section \ref{sec:Bmode_rand} by comparing the B-mode maps from real data with those generated from our mocks. More details about the method, results, and simulations used to perform these tests are described below.

\subsection{Photometric redshift uncertainties}
\label{subsec:photoz_sims}
In our baseline analysis, we define the tomographic redshift bins based on the best estimation by the \texttt{MLZ} algorithm. To assess the impact of photo-$z$ uncertainties on our results, we generate 100 additional \hsc\  realizations of the fiducial model by adopting two alternative redshift estimates: one provided by the \texttt{Mizuki} and the other by the \texttt{Frankenz} method \citep{tanaka2018photometric}. For each of these two mocks, we compute the statistics and calculate their averages, which are then used as the data vector for our cosmology constraints. During the parameter inference, we utilize the emulator constructed based on the \texttt{MLZ} code. Given that the redshift distributions are slightly different among these methods, this test allows us to investigate and quantify any potential biases or inconsistencies that may arise from unaccounted effects associated with photo-$z$ uncertainties. 

We found that the \texttt{Frankenz} model causes a shift in $S_8$ towards negative values of $0.22\sigma$ with respect to the fiducial model, while the \texttt{Mizuki} model shifts it by $0.05\sigma$. Therefore, we do not expect bias larger than the threshold established ($0.3\sigma$) in our analysis caused by different photometric redshift distributions.

\subsection{Image calibration uncertainties}
\label{subsec:calib_sims}
Our mock data are generated with a single multiplicative bias value for each field and tomographic bin. However, the \hsc\ multiplicative bias is determined by image simulations and has an associated uncertainty at the percent level. To explore the impact of this uncertainty, we create two sets of mock data with the multiplicative bias intentionally miscalibrated by $\Delta m = \pm 0.01$. Each set consists of 100 realizations and is generated by modifying Eqs.~\eqref{eq:e1mock} and \eqref{eq:e2mock} as $\boldsymbol{\gamma} \rightarrow (1 + m_{\rm tot} \pm \Delta m)\boldsymbol{\gamma}$, but the mock shear catalogs are intentionally miscalibrated using only $(1 + m_{\rm tot})$ in Eq.~\eqref{eq:shear_2gamma}. We assess the impact of the multiplicative bias uncertainty on our parameter estimation, by employing the average of statistics of the miscalibrated mocks as the data vector. However, we utilize the emulator constructed with well-calibrated mocks to predict the theoretical values.

We found a $\Delta m = + 0.01$  in our maps can bias the $S_8$ by $0.18\sigma$, and by $0.17\sigma$ towards negative values for $\Delta m = -0.01$, both of which meet our criterion.
 
\subsection{Baryon feedback}
\label{sec:baryons}
Baryonic effects significantly suppress the matter power spectrum on small scales \citep{martinet2021impact,ferlito2023millenniumtng}. To investigate the impact of baryons on our analysis and establish the framework for our study, we utilize the $\kappa$TNG dataset \citep{osato2021kappatng}. This dataset is based on the cosmological hydrodynamic simulations IllustrisTNG \citep{pillepich2018first,nelson2019illustristng}. It comprises 10,000 realizations for a set including baryons ($\kappa^{\rm baryons}$) and dark matter only ($\kappa^{\rm DM}$). We weigh the $\kappa$TNG light cones by the \hsc\ redshift distribution to place the source galaxies like our data. Next, we add shape noise with variance 
\begin{equation}
    \sigma^2 = \frac{\sigma_{e}^2}{n_{g}^{\rm eff,1} A_{\rm pix}}\,,
\label{eq:noise}
\end{equation}
where $\sigma_{e}$ is mean intrinsic ellipticity of galaxies ($\sigma_{e} \sim 0.28$), $n_{g}^{\rm eff,1}$ is the galaxy number density in Table~\ref{tab:bins_properties}, and $A_{\rm pix}$ is the solid angle of a pixel. To investigate the scale at which the baryonic effects can be sufficiently mitigated, we apply a Gaussian smoothing filter to the maps following the same procedure as applied to the data (Eq.~\ref{eq:smooth}). 

We compute the power spectrum, peak, and minimum of the $\kappa^{\rm DM}$ and $\kappa^{\rm baryons}$ sets. We then employ the statistics derived from the fiducial cosmology, scaled by the ratio between the hydro-simulations and the dark-matter-only results, as the data vector for parameter constraints. This allows us to quantify the shift in the $\Omega_m- S_8$ plane due to baryons under various scenarios and scale cuts. 

Considering the scale cuts and smoothing scales of our baseline analysis, we found the impact of baryons on $S_8$ is smaller than our predefined acceptable criterion, by $0.15\sigma$.

 %  We smooth the maps to remove small-scale fluctuations. 
 % stringent scale cuts are applied to ensure that the results are not sensitive to baryon modeling. A further study of the impact of Baryons in these NG statistics of the HSC data is explored in \citep{daniela_baryons}

\subsection{Galaxy intrinsic alignment}
\label{subsec:IA}
The intrinsic alignment (IA) of galaxies can lead to systematic errors in weak gravitational lensing measurements, as it introduces additional correlations in the observed shapes of galaxies. 

In the context of the non-linear tidal alignment model (NLA), the strength of this effect is determined by the coupling parameter $A_{\rm IA}$ \citep{bridle2007dark}. Previous 2-pt analyses with \hsc\ data report $A_{\rm IA}= 0.38 \pm 0.70$ \citep{hikage2019cosmology}, and $A_{\rm IA}= 0.91^{+0.27}_{-0.32}$ \citep{hamana_hsc}, also consistent with results using Convolutional Neural Network ($A_{\rm IA}= 0.20^{+0.55}_{-0.58}$) ~\citep{lu2023cosmological}. To study the impact of the intrinsic alignment of galaxies in our statistics, we use the simulations presented in \cite{harnois2022cosmic} that contain IA physically consistent with the NLA model. To match the photometric redshift distribution in our analysis, we weigh these light cones to match the HSC photometric redshift distribution. 

Based on 2-pt HSC constraints on $A_{\rm IA}$, we explore the scenario with minimum and maximum values given $\pm 1\sigma$ of these results, i.e.,  $A_{\rm IA} = 1.18$ and $A_{\rm IA} = -0.32$. Following the procedure described in \cite{harnois2022cosmic}, we multiply the $\epsilon^{\rm IA}_{1/2}$ catalog entries by the $A_{\rm IA}$ amplitudes and combine with the $\gamma_{1/2}$ columns. We refer the readers to \cite{harnois2022cosmic} for more details on the IA mocks. We introduce shape noise to the simulations following Eq.~\eqref{eq:noise}, similar to the procedure used for accounting for baryonic effects.\

Subsequently, similarly to the approach described in Section~\ref{sec:baryons}, we contaminate the synthetic data vector at the fiducial cosmology by multiplying it with the ratio between the results of maps with $A_{\rm IA}$ and $A_{\rm IA}=0$. This allows us to study the impact of intrinsic alignment on our parameter constraints and understand how it influences our cosmological analysis.

For $A_{\rm IA} = 1.18$, $S_8$ is biased by $0.16\sigma$ towards negative values, and by $0.05\sigma$ towards positive values for $A_{\rm IA} = -0.32$, which we deem acceptable for this analysis. It is important to note that our study does not incorporate cross-redshift bins for any of the probes. While this inclusion would likely enhance the constraining power \citep{harnois2021cosmic,zurcher2022dark}, these cross-redshift bins are more susceptible to the impact of IA \citep{harnois2022cosmic}, which would likely require full modelling within the emulator.

\subsection{B-mode maps}
\label{sec:Bmode_rand}

As a null test, we compare the statistics of the B-mode signals obtained from the fiducial simulations with those derived from real data. We expect that the gravitational lensing effect does not produce a divergence-free component, so both the fiducial simulations and real data B-mode maps should primarily contain pure shape noise signals. By comparing these two sets, we can verify if they exhibit consistent signals, which further strengthens our confidence in the accuracy of the noise levels of our simulations and the fidelity of our data.

To estimate the $\chi^2$ and corresponding probability-to-exceed (PTE) between the B-mode signal observed in real data and simulations, we utilize the covariance derived from the B-modes of the fiducial simulations. The results for each scenario are presented in Table~\ref{tab:null_bmodes}. We impose a requirement of PTE $> 1\%$ for the test
to pass. We find consistency with the B-mode results for all statistics and redshift bins.

\begin{table}%[h]
\begin{tabular}{cccc}
\hline
\textbf{Statistics} & \textbf{Photo-z bin} & $\boldsymbol{\chi}^2_{\rm null}$/\rm{dof} & \textbf{PTE ($\%$)} \\ \hline
\
                      & $0.3 < z< 0.6$                        & 9.09/4                                                & 5.90                                 \\ 
                      Power Spectrum                                      & $0.6 < z< 0.9$                        & 4.06/4                                                & 40.51                                \\
                                     & $0.9 < z< 1.2$                        & 3.07/4                                                & 54.52                                \\ \hline
\
                         & $0.3 < z< 0.6$                        & 11.16/13                                              & 59.73                                \\
Peak Counts                                      & $0.6 < z< 0.9$                        & 21.28/13                                              & 6.73                                 \\
                                     & $0.9 < z< 1.2$                        & 12.16/13                                              & 56.31                                \\ \hline
\
                       & $0.3 < z< 0.6$                        & 5.14/9                                                & 82.18                                \\
\multicolumn{1}{l}{}              Minimum Counts   & $0.6 < z< 0.9$                        & 4.93/9                                                & 84.03                                \\
\multicolumn{1}{l}{}                 & $0.9 < z< 1.2$                        & 7.80/9                                                & 55.37                                \\ \hline
\end{tabular}
\caption{Summary of $\chi^2$ per degree of freedom and the probability-to-exceed (PTE) for the null-test using B-modes of our mocks and real data.
}
\label{tab:null_bmodes}
\end{table}

\section{Results}
\label{sec:results}
In this section, we present results from weak lensing statistics of the \hsc\ data, including the angular power spectrum, peak counts, and minimum counts. The tomographic bins, scale cuts, and smoothing scales we adopted in our baseline analysis are described in Section~\ref{sec:method}. 
\subsection{Cosmological Constraints}
In this work, we use the angular power spectrum, peak, and minimum counts of the \hsc\ data to constrain the matter density parameter $\Omega_m$ and the combination sensitive to the structure growth, $S_8 = \sigma_8\sqrt{\Omega_m/0.3}$. 

\begin{figure}%[h]
    \centering
    \includegraphics[scale=0.8]{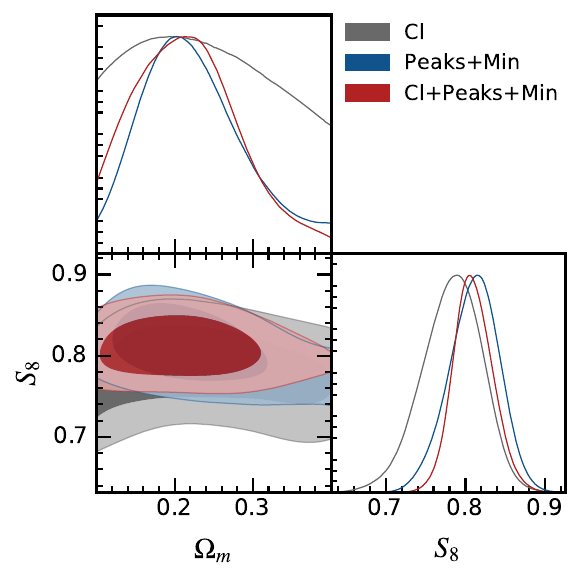}
    \caption{Constraints on the $\Omega_m$ and $S_8\equiv \sigma_8\sqrt{\Omega_m/0.3}$ inferred using the angular power spectra (grey contour) and the joint peak and minimum counts estimation (blue contour) of the \hsc\ data. The red contour is the joint estimation between the three summary statistics.}
    \label{fig:main_constrain}
\end{figure}

In Fig.~\ref{fig:main_constrain}, we show the parameter constraints from the measured statistics in our baseline analysis. In particular, using the angular power spectrum alone (grey contour) we find
\begin{itemize}
\centering
\item[] $\Omega_m = 0.241^{+0.059}_{-0.130} $\,;
\item[] $S_{8} = 0.783^{+0.040}_{-0.034}$\,.
\end{itemize}
Here, we report the 1D marginalized mode and the $\pm 34\%$ confidence intervals. %The best-fit model has $\chi^2 = 6.5$ for 4 dof, corresponding to a PTE 16.4$\%$. 
The constraints from the combination of peak and minimum counts (blue contour of Fig.~\ref{fig:main_constrain}) yield
\begin{itemize}
\centering
\item[] $\Omega_m = 0.225^{+0.048}_{-0.079} $\,;
\item [] $S_{8} = 0.811^{+0.033}_{-0.029}$\,.
\end{itemize}
%The best-fit model for this combination has $\chi^2 = 1.8$ for 2 dof, corresponding to a PTE 40.6$\%$. 

We observe agreement for the results using the different summary statistics, although the results using NG statistics prefer slightly higher $S_8$ values than the power spectrum. 
The constraints of peaks and minima lead to a tightening of the $S_8$ and $\Omega_m$ constraints by approximately $16\%$ and $32\%$, respectively, compared to the constraints obtained from the power spectrum alone.

Combining peak counts, minimum counts, and power spectrum we find

\begin{itemize}
    \centering
    \item[] $\Omega_m = 0.218^{+0.053}_{-0.079} $\,;
    \item [] $S_{8} = 0.810^{+0.022}_{-0.026}$\,,
\end{itemize}
with contour presented in red in Fig.~\ref{fig:main_constrain}. %The best-fit model for this combination has $\chi^2 = 1.7$ for 2 dof, corresponding to a PTE 42.7$\%$. 
This joint constraint represents an improvement of $30\%$ in $\Omega_m$ and $35\%$ in $S_8$ compared to the constraints from the angular power spectrum alone. This improvement highlights the valuable contribution of peak and minimum counts in enhancing the precision of cosmological parameter constraints, which aligns with the findings from other studies \citep{liu2015cosmology, harnois2021cosmic,zurcher2022dark}. 

\begin{figure}%[h]
    \centering
    \includegraphics[scale=0.8]{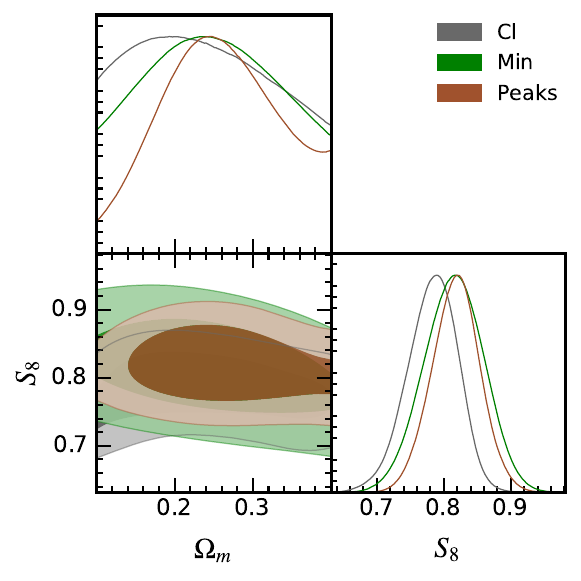}
    \caption{Cosmological constraints (1 and  $2\,\sigma$ contours) obtained from individual summary statistics of \hsc\ data: angular power spectrum (grey), Minimum counts (green), and Peak counts (brown). }
    \label{fig:ind_constrain}
\end{figure}

The primary improvement of the NG statistics on the power spectrum constraints arises from the peak counts, as shown in Fig.~\ref{fig:ind_constrain}. When utilizing peak counts alone, we obtain $\Omega_m = 0.258^{+0.073}_{-0.082}$ and $S_8 = 0.818 \pm 0.035$. For minimum counts, we find $\Omega_m = 0.248 \pm 0.079$ and $S_8 = 0.815 \pm 0.044$. We summarize our main results with the corresponding PTE for the different data combinations in Table \ref{tab:main_result}.

% LFT increased spacing a bit, copied from
% https://tex.stackexchange.com/questions/31672/column-and-row-padding-in-tables
\bgroup
\def\arraystretch{1.2}
\begin{table}
\begin{tabular}{cccc}
\hline
\textbf{Statistics}  & $\boldsymbol{S_8}$            & $\boldsymbol{\Omega_m}$       & \textbf{PTE {[}$\%${]}} \\ \hline
$C_{\ell}$           & $0.783^{+0.040}_{-0.034}$ & $0.241^{+0.059}_{-0.130}$ & 16.4                    \\
Peaks+Min            & $0.811^{+0.033}_{-0.029}$ & $0.225^{+0.048}_{-0.079}$ & 40.6                    \\
Peaks+Min+$C_{\ell}$ & $0.810^{+0.022}_{-0.026}$ & $0.218^{+0.053}_{-0.079}$ & 42.7                    \\
Peaks                & $0.818 \pm 0.035$         & $0.258^{+0.073}_{-0.082}$ & 44.9                    \\
Min                  & $0.815 \pm 0.044$         & $0.248 \pm 0.079$         & 47.2                    \\ \hline
\end{tabular}
\label{tab:main_result}
\caption{Summary of our cosmological constraints for different combinations of statistics and the corresponding probability-to-exceed (PTE).}
\end{table}
\egroup

Our results show consistency with previous studies that used the \hsc\ 2-point information, $S_8 = 0.780^{+0.030}_{-0.033}$ \citep{hikage2019cosmology}, $S_8= 0.823^{+0.032}_{-0.028}$ \citep{hamana_hsc} and $S_8 = 0.812\pm 0.021$ \citep{longley2023unified}. Recently, \cite{dalal2023hyper} and \cite{li2023hyper} conducted two-point analyses using the first three years of data of HSC \citep{li2022three}, which covers an area increased by $\sim 3$ times the \hsc\ area. The results from the Y3 analyses are in agreement with the \hsc\ (and with our results), although the $S_8$ value is slightly lower: $S_8 = 0.776^{+0.032}_{-0.033}$ and $S_8 = 0.769^{+0.031}_{-0.034}$, respectively. Despite the Y3 data having significantly increased the \hsc\ area, the constraining power in $S_8$ for Y3 is similar to \hsc\ \citep{hikage2019cosmology,hamana_hsc} due to the analysis choices adopted to mitigate biases in the photometric redshift estimates. Our results using NG statistics have demonstrated the potential to yield tighter constraints than the power spectrum alone. With this in mind, a future study using NG statistics for the Y3 data could hold the promise of further improving the precision of cosmological parameter constraints. 

In the context of measurements obtained from the 2-point information of other stage-III weak lensing surveys, our results are in agreement within the $1\sigma$ level (considering the sum of the statistical uncertainties in quadrature) with many studies, including KiDS-1000 \citep{asgari2021kids,loureiro2022kids}, DES-Y3 \citep{secco2022dark,amon2022dark,doux2022dark} and KiDS-1000 plus DES Y3 combined \citep{kids_des_2023}. 

Our results are consistent with other stage-III studies that use beyond the 2-point information to constrain $S_8$, such as the peak counts \citep{zurcher2022dark}, and the second and third moments of weak lensing DES Y3 data~\citep{gatti_chang}. The $S_8$ constraints from KiDS-450 and KiDS-1000 data using deep learning \citep{fluri2019cosmological, fluri2022full}, peak counts \citep{martinet2018kids}, and peak counts at high SNR \citep{shan2018kids} are also consistent with our findings. We also find a good agreement with \cite{lu2023cosmological}, which employs peak counts and convolutional neural networks to constrain cosmology with \hsc\ data, and with \cite{liu2023cosmological}, which selected the peak counts at high SNR $> 3.5$ to derived the constraints using a halo-based theoretical peak model. It is important to note that the analysis choices of these studies differ from ours in many aspects, including the redshift bins, cross-correlation between the bins, data vector binning, scale cuts, simulations, treatment of systematic errors, parameters considered during the inference, and modeling of the matter power spectrum and of the summary statistics.

\begin{figure*}
    \centering
    \includegraphics[scale=0.6]{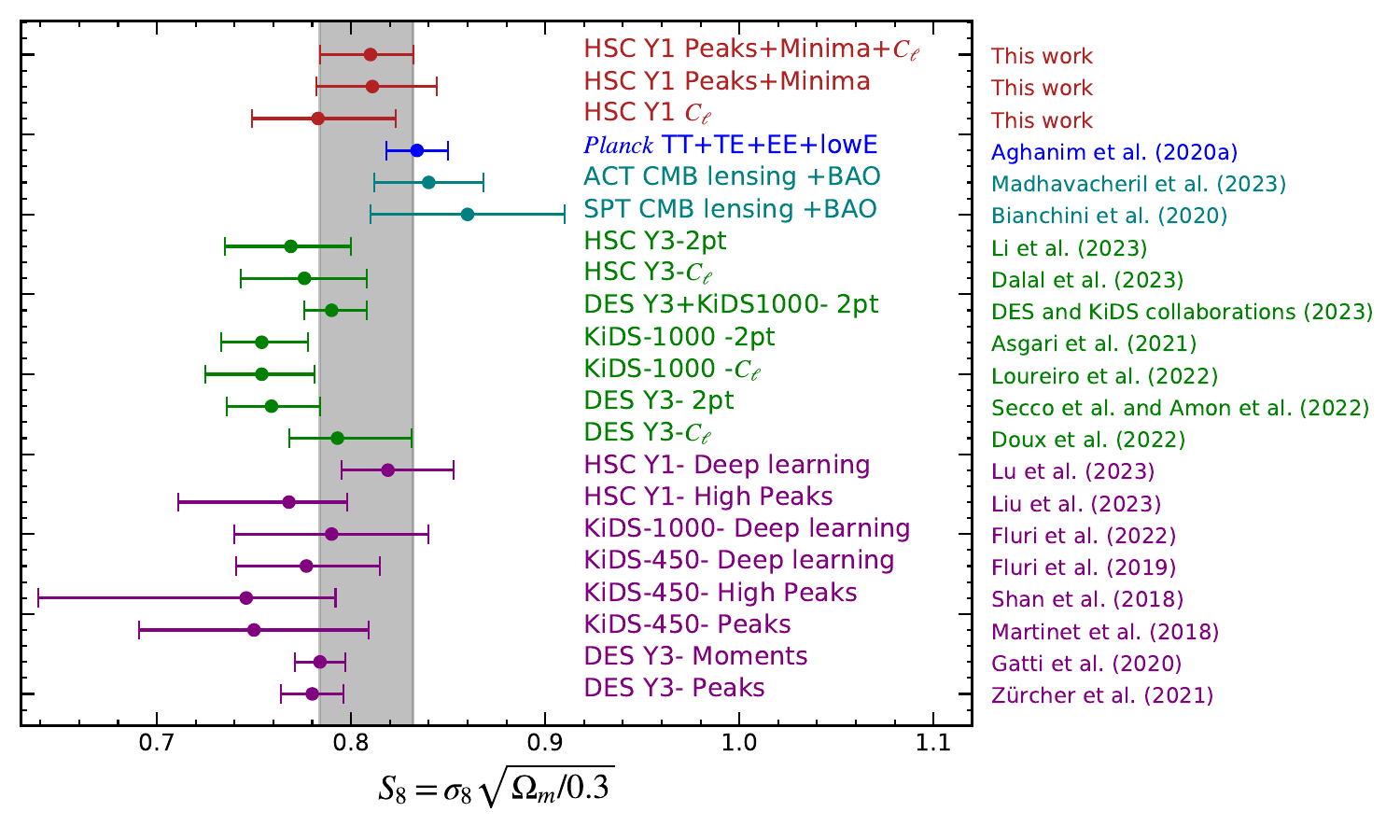}
    \caption{Comparison between our $S_8$ results for different statistical combinations (red points) with external studies: primary CMB (blue point), CMB lensing (teal points), and stage-III weak lensing analyses using 2-point information in real-space (``2pt'') and in harmonic space (``$C_{\ell}$'') (green points). The purple points represent stage-III results from beyond 2-point analyses. We caution the reader that these results stem from distinct datasets as well as different analysis methodologies, including the cosmological and nuisance parameters, priors, model, scale cuts, systematic mitigation strategies, and other specific considerations so that the error bars can vary depending on the setup. For instance, among other differences such as the type of neural network, the results from KiDS-1000 deep learning \citep{fluri2022full} are based on a $w$CDM model, while the KiDS-450 deep learning analysis \citep{fluri2019cosmological} utilizes a $\Lambda$CDM model.}
    
    % We refer the reader to Sec. \ref{sec:results} for more details. }
    %Comparison between our $S_8$ results for different statistical combinations (red points) with other external studies, including: \textit{Planck} primary CMB (blue point) \citep{planckparams}, CMB lensing (teal points) \citep{dr6atacama, qu2023atacama,bianchini2020constraints}, and stage-III cosmic shear analysis using two-point correlation function and power spectrum (green points) \citep{dalal2023hyper,li2023hyper,asgari2021kids,loureiro2022kids,secco2022dark,amon2022dark,doux2022dark}. The purple points represent results from stage-III studies that use beyond 2-point information to constrain $S_8$ \citep{lu2023cosmological, liu2023cosmological,fluri2019cosmological, fluri2022full, shan2018kids,martinet2018kids, gatti_chang, zurcher2022dark}.  }
    \label{fig:summary_literature}
\end{figure*}

The analysis of the TT-TE-EE- and low-E polarization of the \textit{Planck} satellite (\textit{Planck} TT+TE+EE+lowE) found a higher $S_8$ ($S_8=0.834 \pm 0.016$) than the values typically found by cosmic shear analysis \citep{planckparams}. Although our result is slightly lower, we still find a good agreement with the CMB results for all combinations, including the power spectrum, peaks, and minima separately, as well as the power spectrum, peaks, and minimum jointly. We also found good agreement with the results from CMB lensing of the Atacama Cosmology Telescope (ACT) plus Baryon Acoustic Oscillations (BAO) \citep{dr6atacama, qu2023atacama} and from CMB lensing of the South Pole Telescope (SPT) \citep{bianchini2020constraints}. In Figure \ref{fig:summary_literature}, we summarize this comparison of our $S_8$ constraints (red) with the results from external stage-III cosmic shear (green), primary CMB (blue), CMB lensing (teal), and stage-III results from beyond 2-point analyses (purple).

\subsection{Internal consistency checks}
\label{sec:int_consistency}
In previous studies, \hsc\ data have been subjected to comprehensive tests to assess various systematic and astrophysical effects \citep{mandelbaum2018first, oguri2018two,hikage2019cosmology,hamana_hsc,marques2020cross}. In addition to these previous investigations, we have examined the potential impact of the most concerning effects in our results and implemented the necessary analysis choices to mitigate them. To further ensure the robustness of our results, we perform parameter constraints under the following setups:
  \begin{itemize}    
    \item Tomographic bins: We analyze the data using each individual tomographic bin to assess the sensitivity of our results at each redshift. For this test, we jointly consider the power spectrum, minimum, and peak counts to carry out the parameter constraints. 
    \item Removing each tomographic bin: Each individual redshift bin is removed from the analysis to evaluate the influence on the final cosmological constraints. For this test, we jointly consider the power spectrum, minimum, and peak counts to carry out the parameter constraints. 
    \item Smoothing scales: We explore the effect of scales on our NG statistics by performing inference with maps smoothed with the various smoothing scales, including the 4, 5, 8, and $10 \, \mathrm{arcmin}$. For this test, we consider the combination of peak and minimum counts. 
    \item Power spectrum with different scale cuts: We investigate the consequences of varying scale cuts on the angular power spectrum (Cl) to determine the sensitivity of our results to these choices. For this test, we only consider the power spectrum measurements. 
\end{itemize}

\begin{figure}%[h]
    \centering
    \includegraphics[scale=0.6]{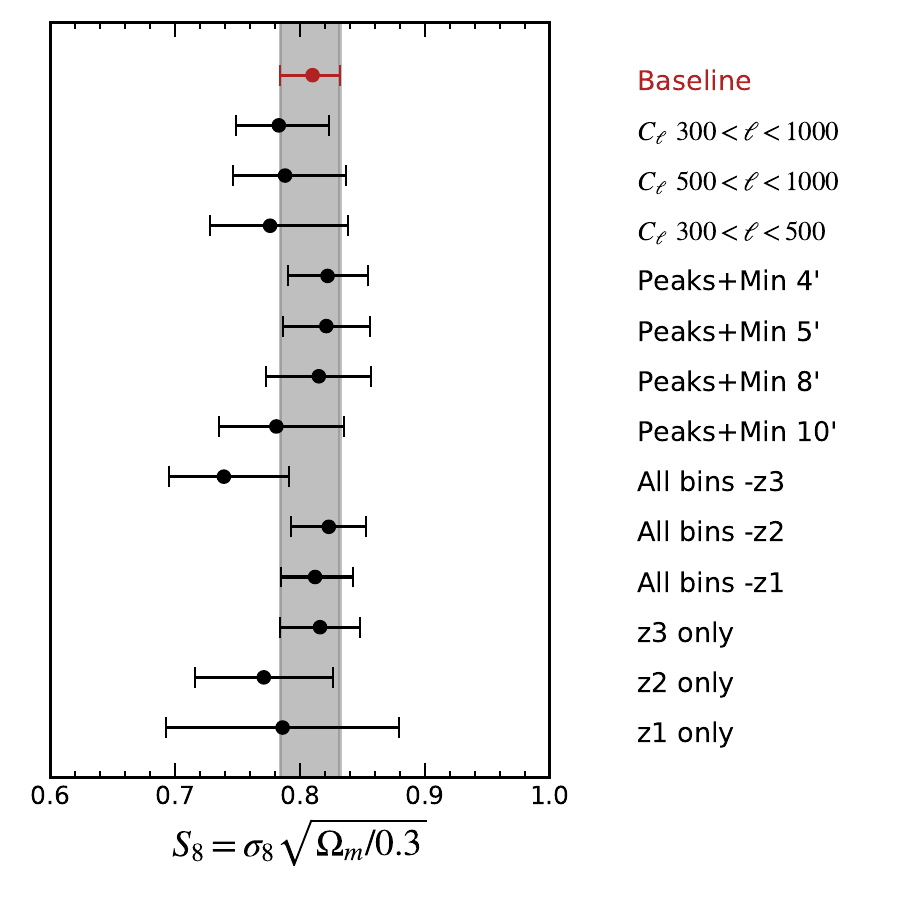}
    \caption{Constraints on $S_8$ and its robustness against various analysis choices described in Section~\ref{sec:int_consistency}. The red point highlights the results from the baseline analysis, where we combine the angular power spectrum, peak counts, and minimum counts of the 3 tomographic bins, along with a scale cut of $300< \ell <1000$ for the power spectrum and $4$ and $8$ $\mathrm{arcmin}$ for the NG statistics. The shaded grey area represents the $68\%$ credible interval for the baseline constraints.
}
    \label{fig:internal_consis}
\end{figure}

In Fig.~\ref{fig:internal_consis}, we show the constraints on $S_8$ for the scenarios described above. We also show in red the results from our baseline setup, that is the angular power spectrum, peak, and minimum counts of the 3 tomographic bins combined, considering the scale cut of $300< \ell <1000$ for the power spectrum and $4 \, \mathrm{arcmin}$ and $8 \, \mathrm{arcmin}$ for the NG statistics. We find consistency within $1\sigma$ in all the different analysis choices, indicating the robustness of our results.

\section{Conclusions}
\label{sec:conclusions}
In this work, we used two non-Gaussian statistics---the peak counts and minimum counts---in addition to the conventional power spectrum, to constrain the matter density $\Omega_m$ and the structure growth parameter $S_8$, assuming a flat-$\Lambda$CDM model. We found that the NG statistics, particularly the peak and minimum counts jointly, significantly tighten the cosmological constraints. %These statistics provide complementary information to the power spectrum, enabling us to achieve tighter constraints on $\Omega_m$ and $S_8$. 
Combining the angular power spectrum, peak counts, and minimum counts, we obtained $68\%$ C.L. constraints of $\Omega_m = 0.218^{+0.053}_{-0.079} $ and $S_{8} = 0.810^{+0.022}_{-0.026}$, an improvement of $30\%$ and $35\%$, respectively, compared to those from the angular power spectrum alone, $\Omega_m = 0.241^{+0.059}_{-0.130} $ and $S_{8} = 0.783^{+0.040}_{-0.034}$. 

To achieve these results, we forward-modelled the summary statistics within our prior range of cosmological parameters, using a suite of $N$-body simulations that incorporate the properties of the \hsc\ data. We tested and validated our pipeline with these simulated mocks prior to analyzing the real data. We assessed the impact of systematics such as intrinsic alignments,
baryon feedback, multiplicative bias, and photometric redshift uncertainties. 
We performed thorough internal consistency tests, finding robust agreements across statistics, angular scales, smoothing scales, and tomographic redshift bins included in our analysis. While the strategy in this work to mitigate systematic effects involved applying scale cuts, smoothing the maps with larger scales, and not including cross-redshift bins for any of the probes, higher-precision studies require addressing the challenges of controlling systematics to robustly constrain cosmology with the information of the non-linear structure of the matter field. We expect future work will improve upon our method by forward-modelling the systematics and marginalizing them without discarding the precious data. 
Our results are in broad agreement with those from previous HSC analyses and with other stage-III weak lensing analyses, using 2-point and non-Gaussian statistics, as summarized in Figure \ref{fig:summary_literature} and discussed in Section~\ref{sec:results}. However, while weak lensing two-point analyses have typically revealed a lower $S_8$ than the ones from CMB, our results are also in good agreement (within 1 $\sigma$) with \textit{Planck} and with CMB lensing results from ACT and SPT \citep{bianchini2020constraints,qu2023atacama,dr6atacama}. 

Our study demonstrated that weak lensing non-Gaussian statistics are not only powerful probes to tighten cosmological constraints beyond 2-point statistics, but also useful tools to investigate the origin of the so-called $S_8$ tension between weak lensing and CMB studies. We expect the constraints to be significantly improved 
%The approach we presented can be a useful asset in future cosmological studies and contribute to the efforts to unravel the growth of structures. An application using ongoing data, such as the HSC Y3 data that covers approximately three times a larger area than Year 1, and 
with the final-year HSC dataset which will cover $\sim$1000  deg$^2$ of the sky. %, can lead to significant improvements in cosmological constraints. 
%
%Looking further ahead, stage-IV weak lensing surveys will play a crucial role in enhancing our understanding of the universe. 
As the deepest large-area weak lensing survey to date, the HSC survey has provided us with a glimpse of what is expected from the forthcoming stage-IV surveys such as \textit{Vera Rubin Observatory LSST}~\citep{LSST:2008ijt}, \textit{Euclid}~\citep{Euclid}, and \textit{Nancy Grace Roman Space Telescope}~\citep{spergel2015wide}. Together, they will probe the growth of structures at unprecedented precision. 
Non-Gaussian statistics, such as the ones studied here, provide excellent opportunities to drive discovery in cosmology with these stage-IV surveys. %While the strategy in this work to mitigate systematic effects, such as baryonic feedback and intrinsic alignments, involved applying scale cuts, smoothing the maps with larger scales, and not including cross-redshift bins for any of the probes, higher-precision studies require addressing the challenges of controlling systematics to robustly constrain cosmology with the information of the non-linear structure of the matter field.  

%Our study represents a step toward better understanding not only how to enhance the constraining power but also comprehending the impact of systematic effects and limitations. 

\section*{Acknowledgements}
We thank Alex Drlica-Wagner, Xiangchong Li, Joaquín Armijo, Sunao Sugiyama, Masahiro Takada, Surhud More, Brian Lu, Zolt\'an Haiman for useful discussions. GM and KMH acknowledge support from the National Science Foundation award 1815887 and the FSU College of Arts and Sciences Dean’s Postdoctoral Scholar Fellows program. This manuscript has been authored by Fermi Research Alliance, LLC under Contract No. DE-AC02-07CH11359 with the U.S. Department of Energy, Office of Science, Office of High Energy Physics. This work was supported by JSPS KAKENHI Grants 23K13095 and 23H00107 (to JL), 	
19K14767 and 20H05861 (to MS), and 21J00011 (to KO). The work of LT is supported by the NSF grant AST~2108078. DG acknowledges support from the ANID-Doctorado Nacional/2019-2119188. JHD acknowledges support from an STFC Ernest Rutherford Fellowship (project reference ST/S004858/1). 
We thank the Yukawa Institute for Theoretical Physics at Kyoto University, where part of the discussions of this work took place during YITP-T-22-03 on ``Cosmology with Weak Lensing: Beyond the Two-point Statistics'' and YITP-W-23-02 on ``Future Science with CMB x LSS''. 
Part of this work was performed at the Aspen Center for Physics, which is supported by National Science Foundation grant PHY-1607611. We acknowledge the use of many public Python packages not cited along the main text: Numpy \citep{oliphant2015guide}, Astropy\footnote{\url{http://www.astropy.org}} a community-developed core Python package for Astronomy~\citep{astropy:2013, astropy:2018}, Matplotlib~\citep{hunter2007matplotlib}, IPython~\citep{perez2007ipython} and Scipy~\citep{jones2001scipy}. 
This research used computing resources at Kavli IPMU. This research used resources at the National Energy Research Scientific Computing Center (NERSC), a U.S. Department of Energy Office of Science User Facility located at Lawrence Berkeley National Laboratory, operated under Contract No. DE-AC02-05CH11231.
The authors are pleased to acknowledge that the work reported in this paper was substantially performed using the Princeton Research Computing resources at Princeton University which is consortium of groups led by the Princeton Institute for Computational Science and Engineering (PICSciE) and Office of Information Technology's Research Computing.\

%%%%%%%%%%%%%%%%%%%%%%%%%%%%%%%%%%%%%%%%%%%%%%%%%%



\bibliographystyle{mnras}
\bibliography{peakcounts} % if your bibtex file is called example.bib

%\begin{thebibliography}{99}
%\end{thebibliography}

%%%%%%%%%%%%%%%%%%%%%%%%%%%%%%%%%%%%%%%%%%%%%%%%%%

%%%%%%%%%%%%%%%%% APPENDICES %%%%%%%%%%%%%%%%%%%%%

\appendix
\newpage
\section{Emulator acccuracy}
\label{apped:emulator_valid}

\begin{figure*}%[h!]
    \centering
    \includegraphics[scale=0.45] {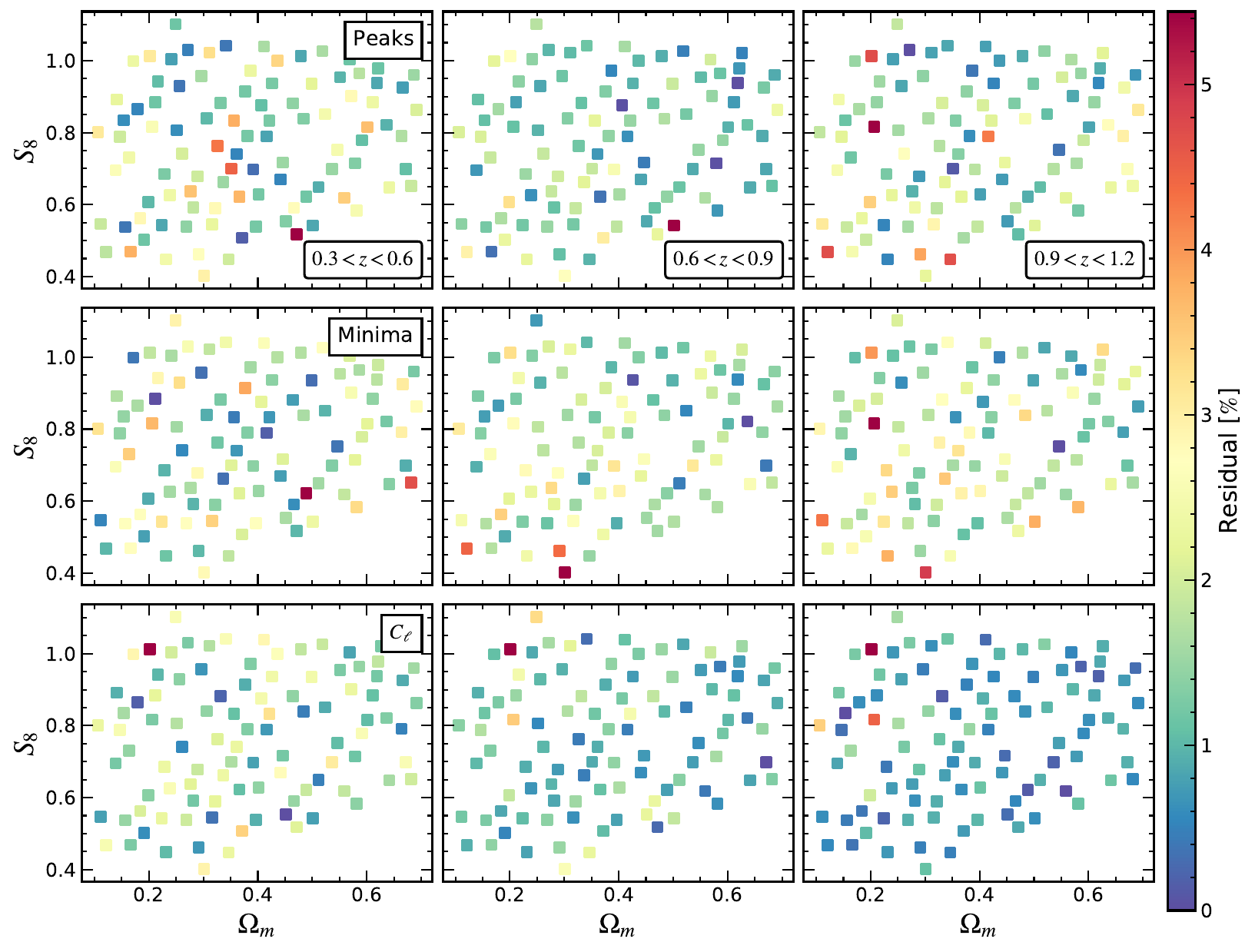}
    \caption{``Leave-one-out'' validation of the Gaussian Process emulator for the Peak Counts (upper panel), minimum counts (middle panel), and Power Spectrum (lower panel) of each tomographic bin and considering maps smoothed with smoothing scale equal to  $4\,\mathrm{arcmin}$. The emulator is built by removing the simulations at each cosmology, and then we compare the prediction with the true data that was removed. The colors denote the amplitude, in percentage, of the mean relative residual.}
\label{fig:emulators}
\end{figure*}

\begin{figure*}%[h!]
    \centering
    \includegraphics[scale=0.45] {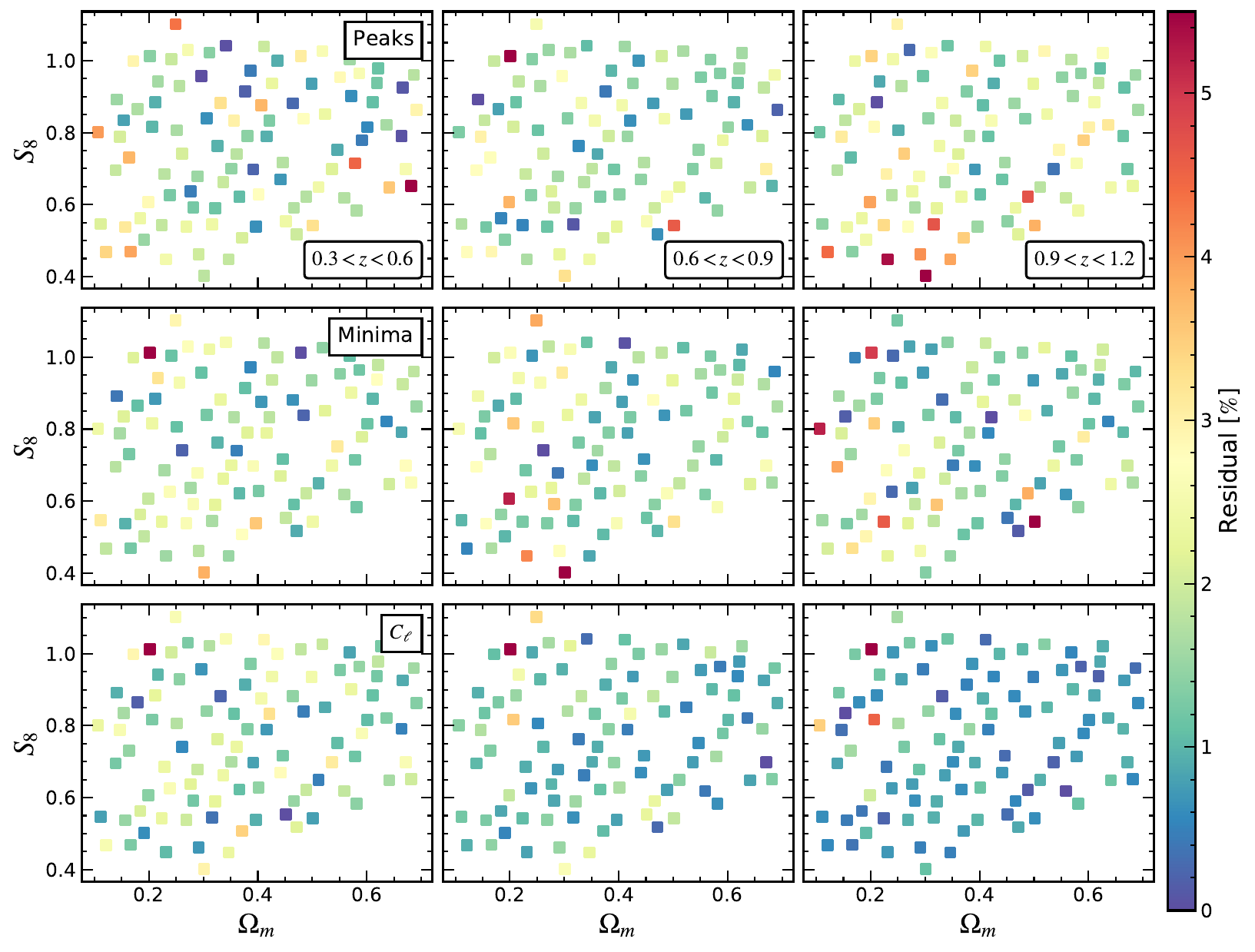}
    \caption{The same as Fig.~\ref{fig:emulators}, but for maps smoothed with smoothing scale equal to $8\,\mathrm{arcmin}$.}
\label{fig:emulators2}
\end{figure*}

We test the accuracy of the emulator by doing the ``leaving-one-out'' validation. We built the emulator by removing each of the 100 cosmologies at each turn and checked how well this removed cosmology can be recovered. Figs.~\ref{fig:emulators} and \ref{fig:emulators2} show the mean residual values for the Peak counts (upper panel), Minima (middle panel), and Power spectrum (lower panel), for maps smoothed with 4 and 8 $\mathrm{arcmin}$, respectively. We found that the relative residual of the emulator does not exceed $\sim 4\%$ within the prior range, enabling unbiased estimation in our analysis.

% Don't change these lines
\bsp	% typesetting comment
\label{lastpage}
\end{document}